\documentclass[12pt]{gtart}
\usepackage{amsmath,amssymb, enumerate,graphicx,epsfig}
\usepackage{accents}

\usepackage[normalem]{ulem}
\usepackage{color,graphicx}

\newcommand{\goesp}{\mbox{ $\stackrel{\rm p}{\longrightarrow}$ }}
\newcommand{\goesd}{\mbox{ $\stackrel{\rm d}{\longrightarrow}$ }}

\def\refhg{\hangindent=25pt\hangafter=1}
\def\refmark{\par\vskip 2mm\noindent\refhg}
\DeclareMathOperator*{\plim}{plim}

\begin{document}

\newtheorem{thm}{Theorem}[section]
\newtheorem{lem}[thm]{Lemma}
\newtheorem*{zlem}{Zorn's Lemma}
\theoremstyle{definition}
\newtheorem{defn}[thm]{Definition}
\newtheorem{rem}[thm]{Remark}
\newtheorem{exam}{Example}[section]
\newtheorem{sub}{Sublemma}[thm]
\newtheorem{prop}[thm]{Proposition}
\newtheorem{cor}[thm]{Corollary}

\begin{center}
\noindent {\bf \Large Quantile correlation coefficient: a new tail dependence measure
	
}

\vspace{10mm}

{\large
\noindent { Ji-Eun Choi and Dong Wan Shin\footnote{Corresponding
		author. Mailing Address: Dept.~of Statistics, Ewha University,
		Seoul, Korea.  Tel: $+$82-2-3277-2614;  Fax: $+$82-2-3277-3606;
		{\it E-mail address}: shindw@ewha.ac.kr (D.W.\ Shin)} }

\noindent {Department of Statistics, Ewha
	University}

\noindent March 16, 2018}
\end{center}

\vspace{10mm}

\noindent {\bf Abstract.} { We  propose a new  measure related with tail dependence in terms of correlation: quantile correlation coefficient of random variables X, Y. The quantile correlation is defined by the geometric mean of two quantile regression slopes of X on Y and Y on X in the same way that the Pearson correlation is related with the regression coefficients of Y on X and X on Y. The degree of tail dependent association in X, Y, if any, is well reflected in the quantile correlation. 
The quantile correlation makes it possible to measure sensitivity of a conditional quantile of a random variable with respect to change of the other variable.
The properties of the quantile correlation are similar to those of the correlation. This enables us to interpret it from the perspective of correlation, on which  tail dependence is reflected. We construct measures for tail dependent correlation and tail asymmetry and develop statistical tests for them.
We prove  asymptotic normality of the estimated quantile correlation and limiting null distributions of the proposed tests, which is well supported in finite samples
by a Monte-Carlo study. The proposed quantile correlation methods are well illustrated by analyzing birth weight data set and stock return data set.}


\vspace{3mm}



\newpage

\section{Introduction}
Correlation coefficient is a standard statistical tool for measuring relationship between two variables. There are several versions of correlation coefficient such as the Pearson correlation coefficient, the Spearman's rank correlation coefficient, the Kendall's tau rank correlation coefficient, and others. The most common of these is the Pearson correlation coefficient, which is a measure for linear association.  However, these correlation coefficients fail to measure tail-specific relationships.

Recently, interests in associations of random variables in tail parts have grown up in various fields.
In finance, recurrent global finance crises have shown that a risky status of one financial institution causes a series of bad impacts on other financial institutions or on the total financial system.  Hence, many studies on the measures for tail dependence have been conducted in the recent literature:
CoVaR (co-value at risk) of Adrian and Brunnermeier (2016) and Giradi and Ergun (2013), volatility spillover index of Diebold and Yilmaz (2012) and many others. Other statistical tools were considered for tail dependence analysis.  Copular is  considered by many authors, see Joe et al. (2010), Nikoloulopoulos et al. (2012) and Kollo et al. (2017). 

In environment, as frequency of abnormal climate has increased, importance for identifying associations of environmental factors in extreme tail part is accentuated. Accordingly, statistical analysis for association between  abnormal climate and other factors using quantile regression have been conducted by many authors: Sayegh et al. (2014) for $PM_{10}$ concentration; Meng and Shen (2014) for extreme temperature; Vilarini et al. (2011) for heavy rainfall and others.

We therefore need a measure which captures tail-specific relations. We define a new correlation coefficient, called  ``quantile correlation coefficient", as a measure related to tail dependence in the context of correlation of random variables $X$ and $Y$.
There is already a measure named ``quantile correlation coefficient", $\rho_{I\tau}^{X,Y}$ say, proposed by Li et al. (2015) which is the Pearson correlation of the indicator $I(X>Q_\tau^X)$ of the event $(X>Q_\tau^X)$ and $Y$ with $\tau$-quantile $Q_\tau^X$ of random variable $X$, $\tau\in(0,1)$. Clearly, $\rho_{I \tau}^{X,Y}$ is not symmetric in $(X,Y)$ in that $\rho_{I \tau}^{X,Y} \ne \rho_{I \tau}^{Y,X}$. Moreover, the measure $\rho_{I\tau}^{X,Y}$ is a compound measure of sensitivity of conditional probability $P(X> Q_\tau^X|Y)$ to change in $Y$ and heterogeneity of conditional expectations $E[Y|X \le Q_\tau^X]$ and $E[Y|X>Q_\tau^X]$, which make it difficult to get a clear interpretation related with tail dependence, see Sections 2, 6.
In fact, $\rho_{I\tau}^{X,Y}$ fails to reflect the degree of tail dependent association as illustrated in Examples 2.1, 2.2 below.
Therefore, it is necessary to define a new quantile correlation coefficient which capture well the degree of tail dependent association and allows a clear interpretation for tail dependence. 

The $\tau$-quantile correlation coefficient $\rho_\tau = sign(\beta_{2.1}(\tau)) \sqrt{\beta_{2.1}(\tau)\beta_{1.2}(\tau)},~0<\tau<1,$ of two random variables $X$, $Y$ is defined by the geometric mean of the two $\tau$-quantile regression slopes $\beta_{2.1}(\tau)$ of X on Y and $\beta_{1.2}(\tau)$ of Y on X. Note that the Pearson  correlation coefficient $\rho = sign(\beta_{2.1}^L)\sqrt{\beta_{2.1}^L\beta_{1.2}^L}$ is the geometric mean of the two linear regression slopes $\beta_{2.1}^L$ of X on Y and $\beta_{1.2}^L$ of Y on X.
The geometric mean $\rho$ indicates overall sensitivity of conditional mean of a variable with respect to change of the other variable. Similarly, the quantile correlation coefficient $\rho_\tau$ has the meaning of overall sensitivity of conditional $\tau$-quantile of one variable with respect to change of the other variable.

Our quantile correlation coefficient will be shown to have many advantages of clear meaning and easy estimation.   The quantile correlation coefficient satisfies the basic features of correlation coefficient: being zero for independent random variables; being $\pm 1$ for perfectly linearly related random variables; commutativity; scale-location-invariance; being bounded by 1 in absolute value for a general class of $(X,Y)$. This allows quantile correlation coefficient to be interpreted as a correlation coefficient. 

The quantile correlation coefficient can be applied diversely. 
First, we can compare how sensitive lower, upper, median conditional quantile of one variable is to unit change of the other variable. For example,  we can identify the fact that a stock return is more affected in lower tail conditional quantiles by change of another stock return than in upper tail conditional quantiles or than in conditional median. 
Second, it can be used to determine the order of variables which have high  sensitivity in tail conditional quantiles with respect to change of the specific variable. For example, in environment, we can use it in primary  screening of environmental factors which cause  abnormal climate, such as high concentration of fine dust, heavy snow, heat wave and many others. 

An estimation method is implemented for the quantile correlation coefficient giving us the sample quantile correlation coefficient. Based on the sample quantile correlation coefficient, we construct new measures and tests for  differences between $\tau$-quantile correlation and the median correlation  and between left $\tau$-quantile correlation and right $(1-\tau)$-quantile correlation. We derive the asymptotic distributions of the sample quantile correlation coefficient and the asymptotic null distributions of the proposed tests. 

A Monte-Carlo experiment shows finite sample validity of asymptotic distribution of the sample quantile correlation coefficient through its stable confidence interval coverage. The experiment also demonstrates that the proposed tests have reasonable sizes and powers. 
The proposed quantile correlation coefficient methods are well demonstrated by analyzing birth weight  data set and stock return data set for investigating the relations between mother's weight $(X)$ gained during pregnancy and  birth weight $(Y)$ and between the US S\&P 500 index return $(X)$ and the French CAC 40 index return $(Y)$.

The remaining of the paper is organized as follows. Section 2 defines quantile correlation coefficient. Section 3 implements an estimation method. Section 4 establishes asymptotic distributions. Section 5 contains a finite sample Monte-Carlo simulation. Section 6 applies the quantile correlation coefficient methods to real data sets. Section 7 gives a conclusion.

\section{Quantile correlation coefficient}
In Section 2.1, quantile correlation coefficient $\rho_\tau$ is defined for a random vector $(X,Y)$ which addresses $\tau$-tail specific relation of $X$ and $Y$, $\tau \in(0,1)$. Meaning of $\rho_\tau$ is discussed to be a sensitivity measure of conditional $\tau$-quantile of a variable with respect to change of the other variable. The proposed $\rho_\tau$ is shown to satisfy the properties what the Pearson correlation coefficient $\rho$ does. In Section 2.2, two examples illustrate that tail-dependent relations of $X$ and $Y$ are well reflected in $\tau$-dependent shape of $\rho_\tau$. Measures of tail-dependency and tail asymmetry are proposed.

\subsection{Definition and properties}\label{sec def}

The quantile correlation is motivated from the relationship between linear regression coefficients and correlation coefficient.  Let $(X,Y)$ be a random vector having finite second moment. Let $\sigma_{XX}=Var(X)$, $\sigma_{YY}=Var(Y)$, $\sigma_{XY}=Cov(X,Y)$.  We observe that $\beta_{2.1}^L=\frac{\sigma_{XY}}{\sigma_{XX}}$ is the $\beta$ minimizing the expected squared error loss $E[(Y-\alpha-X\beta)^2]$ and $\beta_{1.2}^L=\frac{\sigma_{XY}}{\sigma_{YY}}$ is the $\beta$ minimizing $E[(X-\alpha-Y\beta)^2]$. Note that  $\rho=\frac{\sigma_{XY}}{\sqrt{\sigma_{XX}\sigma_{YY}}}=sign(\beta_{2.1}^L)\sqrt{\beta_{2.1}^L\beta_{1.2}^L}$ is  the geometric mean of the two linear regression coefficients. This correlation coefficient $\rho$ measures sensitivity of conditional mean of a random variable with respect to change of the other variable. The correlation is modified to measure sensitivity of conditional quantile rather than of conditional mean by considering $\tau$-quantile regressions of minimizing the expected losses of $\tau$-quantile regression, $0<\tau<1$, rather than linear regressions of minimizing the expected square error losses: the $\tau$-quantile correlation coefficient $\rho_\tau=sign(\beta_{1.2}(\tau))\sqrt{\beta_{1.2}(\tau)\beta_{2.1}(\tau)}$ is defined to be the geometric mean of the two $\tau$-quantile regression coefficients $\beta_{2.1}(\tau)$ and $\beta_{1.2}(\tau)$ of Y on X and X on Y. 

In order to see what $\rho_\tau$ tells us, we first review what the Pearson correlation coefficient $\rho=sign(\beta_{1.2}^L)$ $\sqrt{\beta_{1.2}^L\beta_{2.1}^L}$ tells us. Assume temporarily linearities of $E[Y|X]=\alpha_{2.1}^L+\beta_{2.1}^LX$ and $E[X|Y]=\alpha_{1.2}^L+\beta_{1.2}^LY$.
Note that $\beta_{2.1}^L$ is the amount of change of $E[Y|X]$ with respect to unit change in $X$ and so is $\beta_{1.2}^L$ the amount of change of $E[X|Y]$ with respect to unit change in $Y$. 
The regression coefficients $\beta_{2.1}^L$ and $\beta_{1.2}^L$ are sensitivities of conditional expectations with respect to changes of conditioning variables. When the linearities of conditional expectations are violated, $\beta_{2.1}^L$ and $\beta_{1.2}^L$ are overall sensitivities of changes of conditional expectations with respect to changes of conditioning variables.
Therefore, their geometric mean $\rho$ tells us overall sensitivity of conditional mean of a variable with respect to change of the other variable: the larger $|\rho|$, the more sensitive the conditional mean of one variable to change of the other variable in an overall sense. 
By the same reasoning, the median correlation $\rho_{0.5}$ is an overall sensitivity measure of conditional median of one variable with respect to change of the other variable: the larger $|\rho_{0.5}|$, the more overally sensitive the conditional median of a random variable to change of the other variable. 

Similarly, for given $\tau\in(0,1)$, the larger $|\rho_\tau|$, the more sensitive overally the conditional $\tau$-quantile of a random variable to change of the other variable. 
Therefore, comparison of $\rho_\tau$ for different $\tau$ is meaningful. For example, if $\rho_{0.1}>\rho_{0.5}$, it means that the conditional 0.1-quantile of a random variable is overally more sensitive to change of the other variable than the conditional median of it. If $\rho_{0.1}>\rho_{0.9}$, it means the left conditional 0.1-quantile of a random variable is overally more sensitive to change of the other variable than the right conditional 0.1-quantile of it. Therefore, we can say that
$\rho_\tau$ is an overall sensitivity measure of conditional $\tau$-quantile of one variable with respect to change of the other variable.

On the other hand, the quantile correlation $\rho_{I\tau}^{X,Y}=corr(X^I,Y),~X^I=I(X> Q_\tau^X)$, of Li et al. (2015) has complicated implication, where $Q_\tau^X$ is the $\tau$-quantile of $X$ and $I(A)$ is the indicator function of an event A. Assume linear conditional expectations for $X^I$ and $Y$.  We have 
$E[Y|X^I]=\alpha_{2.1}^I+\beta_{2.1}^IX^I$, $E[X^I|Y]=P(X> Q_\tau^X|Y)=\alpha_{1.2}^I+\beta_{1.2}^IY$, $\rho_{I\tau}^{X,Y}=sign(\beta_{1.2}^I\beta_{2.1}^I)\sqrt{\beta_{1.2}^I\beta_{2.1}^I}$. Note that $\beta_{1.2}^I$ is the change of the conditional probability $P(X>Q_\tau^X|Y)$ associated with unit change in $Y$ and that $\beta_{2.1}^I=E[Y|X> Q_\tau^X]-E[Y|X\le Q_\tau^X]$. Therefore, large $|\rho_{I\tau}^{X,Y}|$ indicates (i) strong sensitivity of the conditional probability of $X > Q_\tau^X$ being highly sensitive to change in $Y$ or (ii) strong heterogeneity of conditional mean of $Y$ having large difference in mean depending on $X > Q_\tau^X$ or $X \le Q_\tau^X$. Therefore, $\rho_{I\tau}^{X,Y}$ is a compound measure of sensitivity of conditional probability $P(X> Q_\tau^X)$ to change in $Y$ and heterogeneity of conditional expectations $E[Y|X > Q_\tau^X]$ and $E[Y|X\le Q_\tau^X]$, see Section 6.1 for a real data illustration.
It is hard to get a simple sensitivity interpretation from $\rho_{I\tau}^{X,Y}$. Moreover, it is obvious that $\rho_{I \tau}^{X,Y}$ lacks symmetry in that $\rho^{X,Y}_{I\tau}\ne \rho_{I \tau}^{Y,X}$. Furthermore, tail dependent association of $X$, $Y$, if any, is not well reflected in $\rho_{I \tau}^{X,Y}$ as illustrated in Examples 2.1, 2.2.
Unlike $\rho_{I \tau}^{X,Y}$, our quantile correlation $\rho_\tau$ has well reflection of the degree of association of $X$, $Y$ as demonstrated in Examples 2.1, 2.2,  has symmetry in $(X,Y)$ and has a clear sensitivity interpretation.

Our $\tau$-quantile regressions of $Y$ on $X$ and $X$ on $Y$ are  defined by minimizing the expected loss, \begin{equation}\label{loss}
L^{X,Y}_\tau(\alpha, \beta) = E[l_\tau(Y-\alpha-\beta X)],~~L_\tau^{Y,X}(\alpha,\beta)=E[l_\tau(X-\alpha-\beta Y)],~~\tau\in(0,1),
\end{equation}
respectively,  where $$l_\tau(e) = e(\tau-I(e<0))$$ is the loss function of the $\tau$-quantile regression.  Let $\tau \in (0,1)$ be given and let \begin{equation}\label{eq1}
(\alpha_{2.1}(\tau), \beta_{2.1}(\tau))=argmin_{\alpha,\beta}L_\tau^{X,Y}(\alpha,\beta),~~(\alpha_{1.2}(\tau),\beta_{1.2}(\tau))=argmin_{\alpha,\beta}L_\tau^{Y,X}(\alpha,\beta).
  \end{equation} Note that these coefficients are more general than the ``usual quantile regression coefficient" which minimizes the conditional loss function $E[L_\tau^{X,Y}(\alpha,\beta)|X]$ under the linearity assumption of the $\tau$-conditional quantiles of $Y$ given $X$, see Koenker (2005, Section 4.1.2). Our quantile regression coefficient is defined without imposing the linearity assumption. If the $\tau$-quantile of Y given X is linear in $X$,  $(\alpha_{2.1}(\tau),\beta_{2.1}(\tau))$ is the same as the ``usual $\tau$-quantile regression coefficients" as shown in Theorem \ref{coef} below.

We define the quantile correlation for a random vector $(X,Y)$ and study its basic properties. 
\begin{defn} \label{def1} 
	\textit{Let $(X,Y)$ be a random vector having finite first order moment. Let $\beta_{2.1}(\tau)$ and $\beta_{1.2}(\tau)$ be defined in (\ref{eq1}). Given $\tau \in (0,1)$, the $\tau$-quantile correlation coefficient $\rho_\tau$ between X and Y is defined as
	$$\rho_\tau^{X,Y} =  sign(\beta_{2.1}(\tau))\sqrt{\beta_{2.1}(\tau) \beta_{1.2}(\tau)}.$$}
\end{defn}  
If relation between $X$ and $Y$ is heterogeneous in that they have different degrees of association depending on left tails of $(X,Y)$, right tails of $(X,Y)$ and other $(X,Y)$, then the heterogeneity is reflected on $\rho_\tau^{X,Y}$. Therefore, $\rho_\tau^{X,Y}$ can be regarded as a tail-dependence measure. This point will be more investigated in Examples 2.1, 2.2, below. 

If $\beta_{2.1}(\tau)\beta_{1.2}(\tau)<0$, $\rho_\tau^{X,Y}$ is not defined. However, the following theorem shows that the proposed quantile correlation is always well-defined. 

\begin{thm}\label{thm welldef}  
	\textit{For all $\tau$, $\beta_{2.1}(\tau)\beta_{1.2}(\tau) \ge 0$.}
\end{thm}

The following theorem states that under the linear quantile function conditions, the quantile regression coefficients are the same as the ``usual quantile regression coefficient" of Koenker (2005, Section 4.1.2) and many others. Let $Q_\tau^Y(X)$ be the conditional $\tau$-quantile of $Y$ given $X$ and let $Q_\tau^X(Y)$ be that of $X$ given $Y$: 
$$P[Y\le Q_\tau^Y(X)|X]=\tau,~P[X\le Q_\tau^X(Y)|Y] = \tau.$$
Note that $Q=Q_\tau^Y(X)$ minimizes the conditional expected loss $E[ l_\tau(Y-Q)|X]$. Similarly, $Q=Q_\tau^X(Y)$ minimizes $E[l_\tau(X-Q)|Y]$. 


\begin{thm}\label{coef}
	 \textit{Assume $Q_\tau^Y(X)$ and $Q_\tau^X(Y)$ are both linear in $X$ and $Y$, respectively, that is,
	 	\begin{equation} \label{quant eq}
	 	 Q_\tau^Y(X)=\alpha_{2.1}^\tau+\beta_{2.1}^\tau X,~~ Q_\tau^X(Y)=\alpha^\tau_{1.2}+\beta_{1.2}^\tau Y
	 	\end{equation}
	 	for some $(\alpha_{2.1}^\tau,\beta_{2.1}^\tau)$ and $(\alpha_{1.2}^\tau, \beta_{1.2}^\tau)$. Then $(\alpha_{2.1}(\tau), \beta_{2.1}(\tau))=(\alpha_{2.1}^\tau,\beta_{2.1}^\tau),~(\alpha_{1.2}(\tau), \beta_{1.2}(\tau))=(\alpha_{1.2}^\tau,\beta_{1.2}^\tau)$.}
\end{thm}

Basic properties of $\rho_\tau$ such as commutativity, scale-location-equivariance, and others are given below.

\begin{thm}\label{thm prop}
\textit{Assume $(X,Y)$ has finite first moment. We have}

\textit{(i) $\rho_\tau^{X,Y} = \rho_\tau^{Y,X}$,}

\textit{(ii) if $a>0, c>0$, $\rho_\tau^{aX+b, cY+d} = \rho_\tau^{X,Y},$}

\textit{(iii) if $Y=\gamma+\delta X$ and $\delta\ne0$, $~\rho_{\tau}^{X,Y}=sign(\delta)$,}

\textit{(iv) if $X$ and $Y$ are independent, $\rho_\tau^{X,Y}=0$.}
\end{thm}

Thanks to Theorem \ref{thm prop} (i), we can write $\rho_\tau^{X,Y}$ by $\rho_\tau$, which will be adopted in the remaining of the paper. According to properties (ii) - (iv), we have $\rho_\tau=\pm 1$ for perfectly linearly related ($X$, $Y$) and $\rho_\tau=0$ for independent $(X,Y)$ and we know that $\rho_\tau$ is invariant under linear transforms of $X$ or of $Y$ with positive slopes.
The following theorem shows that $|\rho_\tau| \le 1$ for a wide class of distributions.  
We therefore can say that $\rho_\tau$ is a correlation measure of $(X,Y)$ for such class. 

\begin{thm} \label{thm 1}
\textit{Assume $(X,Y)$ has finite first moment. We have $|\rho_{\tau}| \le 1$ if  either (i) $\beta_{1.2}(\tau)\le 0$ or (ii) $\beta_{1.2}(\tau)>0$, $(2\tau-1)\Delta_\tau\ge0$, where $\Delta_\tau=E[e_{1.2}^-]E[e_{2.1}^-]-E[e_{1.2}^+]E[e_{2.1}^+]$, $e_{2.1}=Y-\alpha_{2.1}(\tau)-\beta_{2.1}(\tau)X$, $e_{1.2}=X-\alpha_{1.2}(\tau)-\beta_{1.2}(\tau)Y$, $e^+=eI(e \ge 0)$, $e^-=eI(e<0)$. }
\end{thm}

The condition of Theorem \ref{thm 1}  for $|\rho_\tau| \le 1$ needs to be discussed. We have $|\rho_\tau| \le 1$ if $\beta_{1.2}(\tau)\le0$, if $\tau=\frac{1}{2}$, if $\Delta_\tau=0$ or if $(2\tau-1)\Delta_\tau>0$. The first one $\beta_{1.2}(\tau)\le0$ is the case in which conditional $\tau$-quantile of a random variable is negatively associated with the other variable. From the second condition, we have $|\rho_{0.5}|\le 1$ in any case. The third one $\Delta_\tau=0$ is a kind of symmetry of the residuals $e_{1.2}$ and $e_{2.1}$, which is satisfied for the usual symmetric bivariate distributions such as bivariate normal, bivariate t, bivariate uniform and many others. We finally discuss the last condition $(2\tau-1)\Delta_\tau > 0$. For skewed distributions having $\Delta_\tau>0$, we have $|\rho_\tau|\le1$ for $\tau\le \frac{1}{2}$. This is a satisfactory aspect. For the distributions with $\Delta_\tau>0$, left tails of the distributions of $X$, $Y$ are heavier than right tails. Special important such examples are financial asset returns. For such distributions with heavier left tails, people are  more interested in dependence in left tails than in right tails. For distribution having $\Delta_\tau>0$, even though Theorem 2.5 does not guarantee $|\rho_\tau|\le1$ for $\tau>\frac{1}{2}$, it does not mean $|\rho_\tau|>1$ for $\tau >\frac{1}{2}$.

The following theorem characterizes a situation in which the $\tau$-quantile correlation coefficient $\rho_\tau$ is identical with the Pearson correlation coefficient $\rho$. 

\begin{thm}\label{thm norm} 
	\textit{Assume the conditional distribution of Y given X depend on $X$ only through $E(Y|X)$ and $E(Y|X)$ is linear in X. Assume the same one for the conditional distribution of X given Y. Then $\rho_\tau = \rho$ for all $\tau \in(0,1)$.}
\end{thm}

An important special random vector satisfying the conditions of Theorem \ref{thm norm} is the bivariate normal random vector $(X,Y)$ for which we hence have $\rho_\tau=\rho$. Such random vector $(X,Y)$ satisfying the conditions of Theorem \ref{thm norm} has no tail-specific dependence because association between X and Y is exhausted out by the linear conditional expectations $E(Y|X)$ and $E(X|Y)$.

\subsection{Local dependence measure} \label{sec tail}
This subsection starts with a couple of illustrative examples $(X,Y)$ having $\tau$-dependent
quantile correlation coefficient $\rho_\tau$ whose shape reflects the tail-dependent degree of association between $X$ and $Y$. 
Next, it proposes tail dependence measure and of tail asymmetry measure based on $\rho_\tau$. 

\begin{figure}[h]                                        
	\centering                                                
	\includegraphics[width=0.7\textwidth]{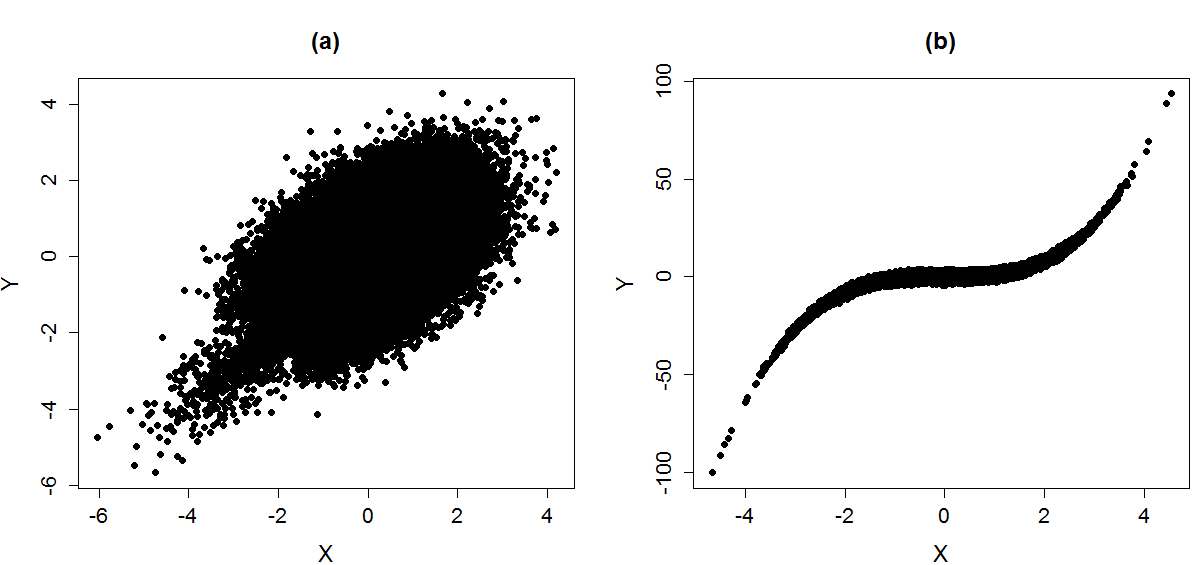}
	\caption{(a) scatter plot of 100000 independent realizations of $(X=\tilde{X}+eI(\tilde{X}<c, \tilde{Y}<c),Y=\tilde{Y}+eI(\tilde{X}<c,\tilde{Y}<c)),~c=-1.645$ in Example 2.1; (b) scatter plot of 100000 independent realizations of $(X,Y=X^3+e)$ in Example 2.2}\label{Fig scatter}
\end{figure}

\begin{exam}\textbf{(A rocket-type bivariate distribution)} Let $(\tilde{X},\tilde{Y}) \sim N_2(0, (\begin{smallmatrix} 1 & \tilde{\rho} \\ \tilde{\rho} & 1 \end{smallmatrix})),~\tilde{\rho}=0.5$. Let $e\sim N(0,1)$ be independent of $(\tilde{X},\tilde{Y})$. Let 
\begin{equation} \label{ex1.2}
X=\tilde{X}+eI(\tilde{X}\le c, \tilde{Y}\le c),~~Y=\tilde{Y}+eI(\tilde{X}\le c, \tilde{Y}\le c),
\end{equation} where $c= -1.645$. Note that the bivariate normal random variables $(\tilde{X},\tilde{Y})$ are contaminated by a common error $e$ in the lower left area of $(X,Y)$. Figure \ref{Fig scatter} (a) displays scatter plot of $(X,Y)$. The figure shows that $(X,Y)$ has a rocket-type scatter plot with stronger correlation in the lower left area $(X\le c, Y\le c)$ than in the other area owing to  the contaminating common term $e$ for X and Y in the lower left area.

Table 1 provides values of quantile correlation coefficient $\rho_\tau$, Pearson correlation $\rho$ for $(X,Y)$ and the quantile correlation coefficients $\rho_{I \tau}^{X,Y}$, $\rho_{I \tau}^{Y,X}$ of Li et al. (2015). We approximate $\rho_\tau$ by a Monte-Carlo simulation average of 100 independent $\hat{\rho}_\tau$ obtained by minimizing the averaged losses $\frac{1}{n}\sum_{i=1}^nl_\tau^{X_i,Y_i}(\alpha,\beta)$ and $\frac{1}{n}\sum_{i=1}^nl_\tau^{Y_i,X_i}(\alpha,\beta)$ with independently generated $(X_i, Y_i),i=1, \cdots, n$. By taking $n$ large enough, by the law of the large numbers, we can make the averaged loss be close enough to the expected loss $L_\tau^{Y,X}(\alpha,\beta)$ and $L_\tau^{X,Y}(\alpha,\beta)$ and hence the approximated value be close enough to the true value $\rho_\tau$. We take $n=1000000$.

  {\footnotesize
  	\begin{center}
  		{\bf Table 1.}  {\color{black}Quantile correlation $\rho_\tau$, $\rho_{I \tau}^{X,Y}$, $\rho_{I \tau}^{Y,X}$ and correlation $\rho$ for $(X,Y)$ in (\ref{ex1.2})}
  		
  		\begin{tabular}{c|ccccccc|c}
  			\hline
  			&&\multicolumn{6}{c }{$\rho_\tau$} \vline& $\rho$\\
  			&$\tau=0.01$ & $\tau=0.05$ & $\tau=0.1$ & $\tau=0.5$ & $\tau=0.9$& $\tau=0.95$  & $\tau=0.99$ &\\ 
  			\hline
  			$\rho_\tau$ &0.553 &0.539 &0.526 &0.499 &0.500 &0.500 &0.500 &0.506\\
  			$\rho_{I \tau}^{X,Y}$  &0.192 &0.231 &0.289 &0.393 &0.292 &0.232 &0.135\\
  			$\rho_{I \tau}^{Y,X}$  &0.192 &0.231 &0.287 &0.397 &0.290 &0.236 &0.135\\
  			\hline
  		\end{tabular}
  	\end{center}
  }
  
  \vspace{2mm}
  
The most interesting point is that $\rho_\tau$ reflects well the degree of association of $X$, $Y$ shown in Figure \ref{Fig scatter} (a). Stronger association of $(X,Y)$ in lower tails of $X$, $Y$ matches up well with larger values of $\rho_\tau$ for $\tau \le 0.1$ than those for $\tau \ge 0.5$, but reversely matches up with smaller values of $\rho_{I \tau}^{X,Y}$ for $\tau \le 0.1$ than that for $\tau =0.5$. Moreover, the fact that $X$, $Y$ have similar degrees of association at center and at high tails by construction is in conflict with different values of $\hat{\rho}^{X,Y}_{I\tau}$ for $\tau=0.5$ and for $\tau$ close to 1. We can say the shape of $\rho_\tau$ is in harmony with the degree of tail-dependent associaiton of $X$, $Y$ while that of $\rho_{I \tau}^{X,Y}$ is not.

The table shows tail dependent features of $\rho_\tau$: the lower quantile correlations  $\rho_{0.01}=0.553$ and $\rho_{0.05}=0.539$ are greater than the median correlation $\rho_{0.5}=0.499$; and the median correlation $\rho_{0.5}=0.499$ is almost the same as the upper quantile correlations $\rho_{0.95}=0.500$ and $\rho_{0.99}=0.500$. 
This is in harmony with the fact that $(X,Y)$ has stronger conditional correlation in the area of (low tail of $X$, low tail of $Y$) than the other area of $(X,Y)$ as observed from Figure \ref{Fig scatter} (a). The fact $\rho_{0.01}=0.553 > \rho_{0.5}=0.499$ tells that, in the overall sense, the 1\% conditional quantile of one random variable varies more sensitively according to change of the other variable than the conditional median. 
This is a consequence of the higher correlation of $(X,Y)$ in the lower left part than in the other part: in the former which, for example, the conditional 0.1-quantile of a variable change overally more sensitively than the conditional median according to change of the other variable because of higher correlation of lower tail $X$, $Y$ than other $X$, $Y$. 
The fact that $\rho_{0.5}\approx\rho_{0.99}= \rho=0.5$ tells that all the conditional median, 99\% quantile, and mean of a random variable vary similarly with change of the other variable. 
\end{exam}

\begin{exam}(\textbf{A cubic relation}) Let $Y= X^3+e$ and let $X$ and $e$ be independent standard normal random variables. Figure \ref{Fig scatter} (b) shows scatter plot of $(X, Y=X^3+e)$. 
We find a tail specific relation between $X$ and $Y$: steeper linear relation at tails than at the center. We provide the values of quantile correlation coefficient $\rho_\tau$ and Pearson correlation coefficient $\rho$ for $(X,Y)$ in Table 2, which are approximated by a Monte-Carlo Simulation, similar to that in Example 2.1.
We note that $\rho_\tau$ has similar tail specific feature as the tail specific relation between $X$ and $Y$: $\rho_\tau$ for tail is larger than $\rho_{0.5}$. This harmonic feature of $\rho_\tau$ and degree of association of $X$,$Y$ is not shared by $\rho_{I \tau}^{X,Y}$, $\rho_{I \tau}^{Y,X}$ in that $\rho_{I \tau}^{Y,X}$ is larger at tails than at center and $\rho_{I,0.01}^{X,Y}<\rho_{I,0.05}^{X,Y}$.

\end{exam} 
{\footnotesize
	\begin{center}
		{\bf Table 2.}  {{\color{black}Quantile correlation $\rho_\tau$, $\rho_{I \tau}^{X,Y}$, $\rho_{I \tau}^{Y,X}$ and correlation $\rho$ for $(X,Y)$, $Y=X^3+e$}}
		
		\begin{tabular}{c|ccccccc|c}
			\hline
			&\multicolumn{7}{c }{$\rho_\tau$} \vline& $\rho$\\
			&$\tau=0.01$ & $\tau=0.05$ & $\tau=0.1$ & $\tau=0.5$ & $\tau=0.9$& $\tau=0.95$  & $\tau=0.99$ &\\ 
			\hline
			$\rho_\tau$ &0.815 &0.745 &0.711 &0.688 &0.711 &0.745 &0.815 &0.750\\
	$\rho_{I \tau}^{X,Y}$  &0.489 &0.560 &0.537 &0.404 &0.541 &0.565 &0.497\\
	$\rho_{I \tau}^{Y,X}$  &0.265 &0.469 &0.559 &0.547 &0.559 &0.470 &0.266\\
			\hline
		\end{tabular}
	\end{center}
}
\vspace{2mm}

The fact that $\rho_{0.5}=0.688<\rho=0.750$ means that, in an overall sense, the conditional median of one variable varies less sensitively with change of the other variable than its conditional mean. From $\rho_{0.01}=\rho_{0.99}=0.815>\rho_{0.5}=0.688,$ we find that the left and right 1\% conditional quantiles of a variable vary more strongly with change of the other variable than its conditional median, which is a consequence of stronger correlation between $X$ and $Y$ in tails than other $(X,Y)$.

The above two examples illustrate that tail-dependent correlation of $(X,Y)$ is well reflected in $\rho_\tau$ but not well in $\rho_{I \tau}^{X,Y}$, $\rho_{I \tau}^{Y,X}$.
Now, we define a new tail dependent correlation measure, which measures how far $\tau$-quantile correlation $\rho_\tau$ is away from the median correlation $\rho_{0.5}$. 

\begin{defn} \textit{Let $\tau \in (0,1)$ be given. The  $\tau$- tail dependent correlation measure $\rho_\tau^D$ is defined as 
	$$ \rho_\tau^D = \rho_\tau -\rho_{0.5}.$$}
\end{defn}
If $\rho_\tau^D \ne0$, we can say that, the conditional $\tau$-quantile of a variable is differently sensitive to change in the other variable than the conditional median of the variable.

\begin{defn}\label{def local} 
	\textit{Tail correlation asymmetry measure is defined as 
	$$ \rho_\tau^A=\rho_{\tau} - \rho_{1-\tau},~~\tau\in(0,0.5).$$}
\end{defn}

It is obvious that the symmetric distributions stated in Theorem \ref{thm norm} having no tail specific correlation have 0 for both of $\rho_\tau^D$ and $\rho_\tau^A$ as formally presented in the following theorem. 

\begin{thm}\label{thm TD=0}
	\textit{For random vector $(X,Y)$ satisfying the conditions of Theorem \ref{thm norm}, $\rho_\tau^D=0$, $\rho_\tau^A=0$ for all $\tau$.}
\end{thm}

\begin{exam}
Consider the $(X,Y)$ in Example 2.1 and Example 2.2. Values of $\rho_\tau^D$ and $\rho_\tau^A$ are computed using $\rho_\tau$ constructed by the Monte-Carlo methods in Examples 2.1, 2.2. Table 3 presents the result.

Consider first the random vectors in Example 2.1. We see tail-dependent values $\rho_\tau^D$: $\rho_\tau^D>0$ for $\tau<0.5$ and  $\rho_\tau^D \cong 0$ for $\tau>0.5$. This means that, compared with the conditional median, as a variable changes, the left tail conditional quantile of the other variable varies more sensitively,
but the right tail conditional quantile varies similarly. We see a highly asymmetric feature by the value $\rho_\tau^A>0$ for $\tau<0.5$.
Consider next $(X,Y)$ in Example 2.2. We observe that $\rho_\tau^D$ and $\rho_{1-\tau}^D$ are the same for $\tau<0.5$, which are larger
for more extreme tails, telling stronger dependency of $(X,Y)$ for deeper tail of $X$, $Y$. We note that $\rho_\tau^A$ has value zero, indicating that $\rho_\tau$ is symmetric and hence symmetric tail dependent relation of $(X,Y)$ in lower tails and in upper tails. 

{\footnotesize
	\begin{center}
		{\bf Table 3.}  {Tail dependent correlation measure ($\rho_\tau^D$), tail correlation asymmetry measure ($\rho_\tau^A$) for Example 2.1 and Example 2.2}
		
		\scalebox{1}{
		\begin{tabular}{c|ccccccc}
			\hline
			$\tau$ &0.01 &0.05 &0.1 &0.5 &0.9 &0.95 &0.99 \\ 
			\hline
	 \multicolumn{1}{l}{Example 2.1} &\multicolumn{7}{l}{$(X,Y)=(\tilde{X},\tilde{Y})+(e,e)I(\tilde{X} \le -1.645, \tilde{Y} \le -1.645)$, }\\
		   \multicolumn{1}{r}{ } & \multicolumn{7}{l}{$e\sim N(0,1),~(\tilde{X},\tilde{Y})\sim N_2(0,(\begin{smallmatrix} 1 & 0.5 \\ 0.5 &1 \end{smallmatrix}))$}\\

			$\rho_\tau^D$  &0.05 &0.04 &0.03 &0.00 &0.00 &0.00 &0.00\\
			$\rho_\tau^A$  &0.05 &0.04 &0.03  &- &- &- &-\\
			\hline
			  \multicolumn{1}{l}{Example 2.2} &\multicolumn{7}{l}{$Y=X^3+e,~(X,e) \sim N_2(0, (\begin{smallmatrix}1 & 0\\ 0 & 1\end{smallmatrix}))$}\\	
			$\rho_\tau^D$ &0.13 &0.06 &0.02 &0.00 &0.02 &0.06 &0.13\\
			$\rho_\tau^A$  &0.00 &0.00 &0.00  &- &- &- &-\\
			\hline
		\end{tabular}}
	\end{center}
}

\vspace{2mm}
\end{exam}

\section{Estimation}\label{sec est}
This section constructs an estimator $\hat{\rho}_\tau$ of $\rho_\tau$ from the sample quantile regressions,  which in turn gives us estimators of $\rho_\tau^D$ and $\rho_\tau^A$. Standard errors for them are presented here, whose theoretical validation is provided in Section \ref{sec asym} and finite sample validation is given in Section \ref{sec simul}. 

Let $X$, $Y$ be two random variables whose tail dependence is of interest.
Suppose that a sample $\{(X_i,Y_i),~i=1,\cdots, n\}$ of $n$ realizations of $(X,Y)$  is given. The sample may be an iid (independent and identically distributed) one or may be possibly non-iid one having linear quantile functions 
\begin{equation} \label{non iid}
Q_\tau^{Y_i}(X_i)=\alpha_{2.1}(\tau)+\beta_{2.1}(\tau)X_i,~~ Q_\tau^{X_i}(Y_i)=\alpha_{1.2}(\tau)+\beta_{1.2}(\tau)Y_i.
\end{equation} 
Note that, for iid samples, we do not need the linearity assumption of (\ref{non iid}). Let $\tau\in(0,1)$ be given. The $\tau$-quantile correlation $\rho_\tau$ is estimated from the sample quantile correlation coefficients as given by
{\begin{equation}\label{est corr}
    	\hat{\rho}_\tau = \begin{cases} sign(\hat{\beta}_{2.1}(\tau))\sqrt{\hat{\beta}_{2.1}(\tau)\hat{\beta}_{1.2}(\tau)},
    	& \text{if $\hat{\beta}_{2.1}(\tau) \hat{\beta}_{1.2}(\tau)\ge0$},\\
    	0,& \text{otherwise},\\
    	\end{cases}
\end{equation}
 
where $(\hat{\beta}_{2.1}(\tau), \hat{\beta}_{1.2}(\tau))$ are the estimated  quantile regression slopes of ($Y$ on $X$, $X$ on $Y$) obtained by minimizing the averaged losses $$\hat{L}_\tau^{X,Y}(\alpha,\beta)=\frac{1}{n}\sum_{i=1}^n l_\tau(Y_i-\alpha-\beta X_i),~~\hat{L}_\tau^{Y,X}(\alpha,\beta)=\frac{1}{n}\sum_{i=1}^n l_\tau(X_i-\alpha-\beta Y_i),$$ that is, 
\begin{equation}\label{eq rhohat}
(\hat{\alpha}_{2.1}(\tau), \hat{\beta}_{2.1}(\tau)) = argmin_{(\alpha,\beta)}\hat{L}_\tau ^{X,Y}(\alpha,\beta),~(\hat{\alpha}_{1.2}(\tau),\hat{\beta}_{1.2}(\tau))=argmin_{(\alpha,\beta)}\hat{L}^{Y,X}_\tau(\alpha,\beta),
\end{equation}
which are the same as the usual $\tau$-quantile regression coefficient estimates. 
The minimization is usually performed  by linear programming, see Koenker (2005, p.181). The estimator $\hat{\rho}_\tau$ in (\ref{est corr}) will be termed as the sample $\tau$-quantile correlation coefficient. The sample quantile correlation can be easily computed from estimated quantile regression estimates using any statistical software for quantile regression estimation such as ``quantreg" in R package, ``proc quantreg" in SAS, and others. For iid sample, $\hat{\rho}_\tau$ is consistent for $\rho_\tau$ with $(\beta_{2.1}(\tau),\beta_{1.2}(\tau))$ defined by (\ref{eq1}) and for possibly non-iid sample having linear quantile functions of (\ref{non iid}), $\hat{\rho}_\tau$ is consistent for $\rho_\tau$ with $(\beta_{2.1}(\tau), \beta_{1.2}(\tau))$ defined by (\ref{non iid}), see Theorem 
\ref{thm rhoq} below.

In finite samples, it may happen that $\hat{\beta}_{1.2}(\tau)\hat{\beta}_{2.1}(\tau)<0$ or greater than 1 for some $\tau$.  For large samples, the probability of $\hat{\beta}_{1.2}(\tau)\hat{\beta}_{2.1}(\tau)<0$ goes to 0 as $n\rightarrow \infty$, as will be demonstrated in Corollary \ref{cor beta} in Section \ref{sec asym} below. 

Section \ref{sec asym} below will show that $\hat{\rho}_\tau$ is asymptotically normal, 
\begin{equation} \label{rho conv}
\sqrt{n} (\hat{\rho}_\tau-\rho_\tau) \goesd N(0,V(\rho_\tau))
\end{equation} for $V(\rho_\tau)$ in (\ref{V conv}) below. 
We present the standard error $se(\hat{\rho}_\tau)=\sqrt{\hat{V}(\hat{\rho}_\tau)}$ of $\hat{\rho}_\tau$ based on a consistent estimator $\hat{V}(\rho_\tau)$ of the asymptotic variance $V(\rho_\tau)$. Let $\theta = (\alpha,\beta)'$and $\theta_1$ and $\theta_2$ be two given $\theta$ values.
Let $X_i^*=(1,X_i)'$ and $Y_i^*=(1,Y_i)'$ and let $$D_\tau(X,Y,\theta_1, \theta_2) = [d_\tau'(X, Y, \theta_1)~|~ d_\tau'(Y, X, \theta_2)]' ,~~ d_\tau(X,Y,\theta) = X^*(\tau-I(Y-\theta'X^*<0)).$$
Given two values of $\tau_1,~\tau_2 \in(0,1)$ and  a density functions $f$, we define
\begin{equation}  \label{H}
\begin{aligned}
H_{\tau_1\tau_2}&(\theta_1, \theta_2) = \plim_{n\rightarrow \infty} \frac{1}{n} \sum_{i=1}^n D_{\tau_1}(X_i, Y_i, \theta_1, \theta_2)D_{\tau_2}'(X_i, Y_i, \theta_1, \theta_2), \\
\end{aligned}
\end{equation}
\begin{equation} \label{M}
\begin{aligned}
M(\theta_1,\theta_2) = \plim_{n\rightarrow \infty} \frac{1}{n}&\sum_{i=1}^n\left[\begin{array}{c|c}
R(f_{Y_i,X_i}, X_i, \theta_1) & 0 \\ \hline 0 & R(f_{X_i,Y_i},Y_i,\theta_2)\end{array}
\right],~~
R(f,X,\theta) = f(\theta'X^*)X^*X^{*'},
\end{aligned}
\end{equation}
{\color{black}which is assumed to exist.} 
Then, with $\alpha_{2.1} = \alpha_{2.1}(\tau)$, $\beta_{2.1}=\beta_{2.1}(\tau)$, $\alpha_{1.2}=\alpha_{1.2}(\tau)$, $\beta_{1.2} = \beta_{1.2}(\tau)$, $\theta_{2.1}=(\alpha_{2.1},\beta_{2.1})'$, $\theta_{1.2}=(\alpha_{1.2},\beta_{1.2})',$ as will be shown in Section \ref{sec asym}, we have
\begin{equation}  \label{V conv}
\begin{aligned}
V({\rho}_{\tau}) = G_1'(\beta_{2.1},\beta_{1.2})[M(\theta_{2.1}, \theta_{1.2})]^{-1} H_{\tau\tau}(\theta_{2.1}, \theta_{1.2})
\times[M( \theta_{2.1}, \theta_{1.2})]^{-1}G_1(\beta_{2.1},\beta_{1.2}),
\end{aligned}
\end{equation}
where $G_1(\beta_1, \beta_2) = \frac{1}{2}(0,g(\beta_1,\beta_2),0,g(\beta_2,\beta_1))',$ $g(\beta_1,\beta_2)=sign(\beta_1)\sqrt{\frac{\beta_2}{\beta_1}},$ if $\beta_1\beta_2>0$; $g(\beta_1,\beta_2)=0,$ otherwise, and $f_{Y_i\cdot X_i}, f_{X_i\cdot Y_i}$ are the conditional densities of $Y_i$ given $X_i$ and $X_i$ given $Y_i$, respectively. 

The results (\ref{rho conv})-(\ref{V conv}) enable us to construct a standard error of $\hat{\rho}_\tau$. We need a consistent estimator $\hat{V}(\rho_\tau)$ of the asymptotic variance $V(\rho_\tau)$ for which we need estimators of the parameter values $\theta_{2.1}$, $\theta_{1.2}$ and conditional density values $f_{Y_i\cdot X_i}(\theta_{2.1}'X_i^*)$, $f_{X_i\cdot Y_i}(\theta_{1.2}'Y_i^*)$ evaluated of $\theta_{2.1}'X_i^*$ and $\theta_{1.2}'Y_i^*$. 
We use the consistent estimators $\hat{\theta}_{2.1} = (\hat{\alpha}_{2.1},\hat{\beta}_{2.1})',~\hat{\theta}_{1.2} = (\hat{\alpha}_{1.2}, \hat{\beta}_{1.2})',~\hat{\alpha}_{2.1} = \hat{\alpha}_{2.1}(\tau),~\hat{\beta}_{2.1}=\hat{\beta}_{2.1}(\tau),~\hat{\alpha}_{1.2}=\hat{\alpha}_{1.2}(\tau),~\hat{\beta}_{1.2}=\hat{\beta}_{1.2}(\tau)$ for the corresponding parameter values. Given estimators $\hat{f}_{Y_i\cdot X_i}$ and $\hat{f}_{X_i\cdot Y_i}$ of $f_{Y_i\cdot X_i}$ and $f_{X_i\cdot Y_i}$, the obvious estimator of $V(\rho_\tau)$ is 
\begin{equation} \label{V}
\hat{V}(\rho_\tau) = \hat{G}_1' \hat{M}^{-1} \hat{H}_{\tau\tau} \hat{M}^{-1}
\hat{G}_1,
\end{equation}
where, 
$$\hat{G}_{1} = G_1(\hat{\beta}_{2.1}, \hat{\beta}_{1.2}),~~\hat{H}_{\tau_1\tau_2} = \frac{1}{n} \sum_{i=1}^n D_{\tau_1}(X_i, Y_i, \hat{\theta}_{2.1}, \hat{\theta}_{1.2})D_{\tau_2}'(X_i, Y_i,\hat{\theta}_{2.1}, \hat{\theta}_{1.2}),~\tau_1=\tau_2=\tau$$ and 
\begin{equation}\label{eq M}
\hat{M}= \frac{1}{n} \sum_{i=1}^n \left[\begin{array}{c|c}
R(\hat{f}_{Y_i\cdot X_i}, X_i, \hat{\theta}_{2.1}) & 0 \\ \hline 0 & R(\hat{f}_{X_i\cdot Y_i},Y_i,\hat{\theta}_{1.2})\end{array}
\right].
\end{equation}
We now have the standard error of $\hat{\rho}_\tau$, $se(\hat{\rho}_\tau)=\sqrt{n^{-1}\hat{V}(\rho_\tau)}.$

Note that estimators of the conditional densities $f_{Y_i\cdot X_i}$ and $f_{X_i\cdot Y_i}$ evaluated at $\theta_{2.1}'X_i^*$ and $\theta_{1.2}'Y_i^*$ are required for variance estimator (\ref{V}) but not for the quantile correlation coefficient estimator $\hat{\rho}_\tau$.
For the estimated conditional kernel density values $\hat {f}_{Y_i\cdot X_i}(\hat{\theta}_{2.1}'X_i^*)$ and $\hat {f}_{X_i\cdot Y_i}(\hat{\theta}_{1.2}'Y_i^*)$ in (\ref{eq M}), two strategies are available. The first strategy is using nonparametric conditional density estimators, $\hat{f}_{Y\cdot X}^{BH}$ and $\hat{f}_{X\cdot Y}^{BH}$, say, of Hyndman et al. (1996) for iid sample. Since $(X_i,Y_i),~i=1,\cdots,n$ are iid, we can omit subscript $i$ as in $f_{X\cdot Y}$ and $f_{Y\cdot X}$. The conditional kernel density estimator of Hyndman et al. (1996) is 
\begin{equation} \label{cond dist2}
\hat{f}^{BH}_{Y\cdot X}(y|x) = \sum_{j=1}^n  \omega_j^X(x) \frac{1}{b_{Y\cdot X}} K(\frac{|y-Y_j|}{b_{Y\cdot X}}),~~\hat{f}^{BH}_{X\cdot Y}(x|y) = \sum_{j=1}^n  \omega_j^Y(y) \frac{1}{b_{X\cdot Y}} K(\frac{|x-X_j|}{b_{X\cdot Y}}),
\end{equation}
where $$\omega_j^X(x)= K(\frac{|x-X_j|}{a_{Y\cdot X}})/\sum_{i=1}^nK(\frac{|x-X_i|}{a_{Y\cdot X}}),~~
\omega_j^Y(y)= K(\frac{|y-Y_j|}{a_{X\cdot Y}})/\sum_{i=1}^nK(\frac{|y-Y_i|}{a_{X\cdot Y}}),$$ $a_{X\cdot Y}$, $b_{X\cdot Y}$, $a_{Y\cdot X}$, $b_{Y\cdot X}$ are bandwidths, and
$K(\cdot)$ is a kernel function.  Optimal choices of the bandwidths are given in Bashtannyk and Hyndman (2001), which will be employed in (\ref{ab}) in Section \ref{sec simul} below. The estimator (\ref{cond dist2}) is consistent for iid sample but not for non-iid sample. The other strategy is using the estimated values of the conditional densities at the quantiles, for example by  Hendricks and Koenker (1991). 
For possibly non-iid smaple, we can use the method of Hendricks and Koenker (1991) who estimated the values of the conditional densities at the estimated quantiles 
under the linearity assumption that quantile $Q_\tau^{Y_i}(X_i)$ of $Y_i$ is linear in $X_i$ and so is $Q_\tau^{X_i}(Y_i)$:
\begin{equation}\label{cond dist}
\begin{aligned}
\hat {f}_{Y_i\cdot X_i}^{HK}(\hat{\theta}_{2.1}'X_i^*) = \max \left(0, \frac{2h_n}{\hat{\theta}_{2.1}'(\tau+h_n)X_i^*-\hat{\theta}_{2.1}'(\tau-h_n)X_i^*-\epsilon}\right),\\
\hat {f}_{X_i\cdot Y_i}^{HK}(\hat{\theta}_{1.2}'Y_i^*) =\max\left(0, \frac{2h_n}{\hat{\theta}_{1.2}'(\tau+h_n)Y_i^*-{\color{black}\hat{\theta}_{1.2}'(\tau-h_n)}Y_i^*-\epsilon}\right),
\end{aligned}
\end{equation}
where $\epsilon$ is a small positive constant and $h_n$ is a bandwidth. Optimal bandwidth parameter $h_n$ is found in Bofinger (1975) and Koenker (2005, p.115), which will be adopted in a Monte-Carlo study in (\ref{h}) in Section \ref{sec simul}.
The estimator (\ref{cond dist}) is consistent for some non-iid samples under some regularity conditions such as linearity of quantile functions
$Q_\tau^X(Y)$ and $Q_\tau^Y(X)$. Performances of these two strategies will be compared in Section \ref{sec simul}.

The tail dependent correlation measure $\rho_\tau^D$ and the tail asymmetry measure $\rho_\tau^A$ are estimated by
\begin{equation}\label{TD TA}
\hat{\rho}_\tau^D = \hat{\rho}_\tau-\hat{\rho}_{0.5},~~\hat{\rho}_\tau^A = \hat{\rho}_\tau-\hat{\rho}_{1-\tau}.
\end{equation}
For standard errors of the estimated measures $\hat{\rho}_\tau^D$ and $\hat{\rho}_\tau^A$, we need the asymptotic variance $V(\rho_{\tau_1}-\rho_{\tau_2})$ of $\hat{\rho}_{\tau_1} - \hat{\rho}_{\tau_2}$ in {\color{black} $\sqrt{n}((\hat{\rho}_{\tau_1}-\hat{\rho}_{\tau_2})-({\rho}_{\tau_1}-{\rho}_{\tau_2})) \goesd N(0, V(\rho_{\tau_1}-\rho_{\tau_2}))$, $\tau_1, \tau_2 \in (0,1)$}, as will be demonstrated in Section \ref{sec asym}, where
\begin{equation} \label{var diff}
\begin{aligned}
V({\rho}_{\tau_1}-{\rho}_{\tau_2}) = G_2'(\beta_{2.1}(\tau_1), \beta_{1.2}(\tau_1),&\beta_{2.1}(\tau_2), \beta_{1.2}(\tau_2)) V({M}, H_{\tau_1\tau_1},  H_{\tau_1\tau_2}, H_{\tau_2\tau_1}, H_{\tau_2\tau_2})\\
\times G_2(\beta_{2.1}(\tau_1),& \beta_{1.2}(\tau_1),\beta_{2.1}(\tau_2), \beta_{1.2}(\tau_2)),
\end{aligned}
\end{equation}
 $$V({M}, H_{\tau_1\tau_1},  H_{\tau_1\tau_2}, H_{\tau_2\tau_1}, H_{\tau_2\tau_2}) = V_1^{-1}V_0 V_1^{-1},$$
 $$ V_1 = \left[\begin{array}{c|c}
 M(\theta_{2.1}(\tau_1), \theta_{1.2}(\tau_1)) & 0 \\ \hline 0 & M(\theta_{2.1}(\tau_2), \theta_{1.2}(\tau_2))\end{array}
 \right]$$
$$V_0=\left[\begin{array}{c|c}
	H_{\tau_1\tau_1}(\theta_{2.1}(\tau_1), \theta_{1.2}(\tau_1)) & H_{\tau_1\tau_2}(\theta_{2.1}(\tau_1), \theta_{1.2}(\tau_2)) \\ \hline H_{\tau_2\tau_1}(\theta_{2.1}(\tau_2), \theta_{1.2}(\tau_1))  & H_{\tau_2\tau_2}(\theta_{2.1}(\tau_2), \theta_{1.2}(\tau_2))\end{array} \right],$$
$$G_2(\beta_1, \beta_2, \beta_3, \beta_4) = \frac{1}{2}(0,\sqrt{ \frac{\beta_2}{\beta_1}},0,\sqrt{\frac{\beta_1}{\beta_2}},0,-\sqrt{\frac{\beta_4}{\beta_3}},0,-\sqrt{\frac{\beta_3}{\beta_4}})'.$$

A consistent estimator of $V({\rho}_{\tau_1} - {\rho}_{\tau_2})$ is 
\begin{equation}\label{var diff est} 
\widehat{V}({\rho}_{\tau_1} - {\rho}_{\tau_2})=
\hat{G}_2' \hat{V}\hat{G}_2,
\end{equation}
where {\color{black}$\hat{G}_2 = G_2(\hat{\beta}_{2.1}(\tau_1), \hat{\beta}_{1.2}(\tau_1),\hat{\beta}_{2.1}(\tau_2), \hat{\beta}_{1.2}(\tau_2))$} and $\hat{V} = 
V(\hat{M}, \hat{H}_{\tau_1\tau_1}, \hat{H}_{\tau_1\tau_2}, \hat{H}_{\tau_2\tau_1}, \hat{H}_{\tau_2\tau_2})$.
The standard errors of $\hat{\rho}_\tau^D$ and $\hat{\rho}_\tau^A$ are 
\begin{equation} \label{TD se}
se(\hat{\rho}_\tau^D) = \sqrt{n^{-1} \hat{V}({\rho}_{\tau}-{\rho}_{0.5})},~~ se(\hat{\rho}_\tau^A) = \sqrt{n^{-1} \hat{V}({\rho}_{\tau}-{\rho}_{1-\tau})}.
\end{equation}

\section{Asymptotic theory and statistical inference}\label{sec asym}
Let $n$ realizations $\{(X_i,Y_i),i=1,\cdots,n\}$ of two random variables $X$, $Y$ be given. 
We establish asymptotic normality for $\hat{\rho}_\tau$ and other estimators in Section 3, which enable us to construct confidence intervals and tests for $\rho_\tau$, $\rho_\tau^D$, $\rho_\tau^A$.
Let $\theta_{1.2}(\tau) =(\alpha_{1.2}(\tau),\beta_{1.2}(\tau))'$ and $\theta_{2.1}(\tau) = (\alpha_{2.1}(\tau), \beta_{2.1}(\tau))'$ be the vectors of $\tau$-quantile regression coefficients defined in (\ref{eq1}) or (\ref{non iid})  for iid sample or for possibly non-iid sample, respectively.
The asymptotic distribution of $\hat{\rho}_\tau$ is established under the following conditions.


\noindent\textbf{Condition A1.} We have either  

\noindent (i) $(X_i,Y_i),~i=1,\cdots,n$  are iid having finite second moment or 

\noindent (ii) $(X_i,Y_i),~i=1,\cdots,n$ are possibly non-iid; $Q_\tau^{Y_i}(X_i)$ is linear in $X_i$ and $Q_\tau^{X_i}(Y_i)$ is linear in $Y_i$ as in (\ref{non iid}); $D_{\tau i}=(d_\tau(X_i,Y_i,\theta_{2.1}(\tau)),d_\tau(Y_i,X_i,\theta_{1.2}(\tau)))'$ is a martingale difference with respect to $F_i=\{(X_j,Y_j),j=1,\cdots,i\}$ satisfying
\begin{equation} \label{cond A1 eq}
\begin{aligned}
\frac{1}{n}\sum_{i=1}^n E[D_{\tau_1 i}D_{\tau_2 i}'|F_{i-1}] & \goesp H_{\tau_1\tau_2}(\theta_{2.1},\theta_{1.2}),
\end{aligned}
\end{equation}
\begin{equation} \label{cond A1 eq2}
\begin{aligned}
\frac{1}{n}\sum_{i=1}^n E[R(f_{X_i\cdot Y_i},Y_i, \theta_{1.2})|F_{i-1}] \goesp \bar{R}_{1.2}(\theta_{1.2}),~~ 
\frac{1}{n}\sum_{i=1}^n E[R(f_{Y_i\cdot X_i},X_i,\theta_{2.1})|F_{i-1}] \goesp \bar{R}_{2.1}(\theta_{2.1}).
\end{aligned}
\end{equation}
Note that, for iid sample of condition A1(i), linearity is not assumed for the quantile functions and (\ref{cond A1 eq})-(\ref{cond A1 eq2}) are automatically satisfied with $E[D_{\tau_i}]=0$ as proved in the proof of Lemma 4.1. Thanks to (\ref{cond A1 eq})-(\ref{cond A1 eq2}), the probability limits in (\ref{H})-(\ref{M}) are well defined for true $\theta_1$, $\theta_2$. Sequence $(X_i,Y_i),~i=1,2,\cdots$ of random vectors having GARCH-type conditional heteroscedasticity satisfy the condition of A1(ii), for which the asymptotic results of this section hold. Therefore, our quantile correlation method is applicable for tail-dependent analysis of financial return data sets.

\noindent\textbf{Condition A2.} Let $\tau \in (0,1)$ be given. The conditional distribution function $F_{Y_i\cdot X_i}(y|x)$ of $Y_i$ given $X_i=x$ is absolutely continuous, with continuous conditional density $f_{Y_i\cdot X_i}(y|x)$ uniformly bounded away from 0 and $\infty$ at $y=\alpha_{2.1}(\tau)+x\beta_{2.1}(\tau)$. The conditional distribution function $F_{X_i\cdot Y_i}(x|y)$ of $X_i$ given by $Y_i=y$ satisfies the similar conditions with conditional density $f_{X_i\cdot Y_i}(x|y)$. 

\noindent\textbf{Condition A3.} Let $\tau \in (0,1)$ be given.
 
\begin{flushleft}
	
(i) $E[f_{Y_i\cdot X_i}(\alpha_{2.1}(\tau)+\beta_{2.1}(\tau)X_i)X_i^2]  ~\text{ and }~ E[f_{X_i\cdot Y_i}(\alpha_{1.2}(\tau)+{\color{black}\beta_{1.2}(\tau)}Y_i)Y_i^2],~~i=1,\cdots,n$ are bounded,
\end{flushleft}

(ii) $E[ \max_i |X_i|]=o(\sqrt{n})$ and $E[\max_i|Y_i|]=o(\sqrt{n})$,


Conditions A2-A3 are similar to those assumed in quantile regression asymptotic analysis, see Section 4.2 of Koenker (2005). Thanks to Condition A2, the asymptotic variance of $\hat{\rho}_\tau$ is expressed in terms of the conditional densities through the terms in Condition A3(i).

Let $\Theta(\tau)=(\theta_{2.1}'(\tau),\theta_{1.2}'(\tau))'$.
In Lemma \ref{lem beta}, we first derive the asymptotic distribution of the vectors of estimated quantile regression coefficients $\hat{\Theta}(\tau) = (\hat{\theta}_{2.1}'(\tau),\hat{\theta}_{1.2}'(\tau))' = (\hat{\alpha}_{2.1}(\tau),\hat{\beta}_{2.1}(\tau),$ $\hat{\alpha}_{1.2}(\tau),\hat{\beta}_{1.2}(\tau))'$.

\begin{lem}\label{lem beta} \textit{Let $\tau \in (0,1)$ be given. Assume conditions A1 - A3. Then, as $n\rightarrow \infty$, we have {\small
\begin{equation}\label{eq lem beta}
\sqrt{n}(\hat{\Theta}(\tau) -\Theta(\tau)) \goesd {\color{black}N_4}(0,[M(\theta_{2.1}, \theta_{1.2})]^{-1}H_{\tau\tau}(\theta_{2.1},\theta_{1.2})[M(\theta_{2.1}, \theta_{1.2})]^{-1}),
\end{equation}}
where $\theta_{2.1}=\theta_{2.1}(\tau)$ and $\theta_{1.2}=\theta_{1.2}(\tau)$.}
\end{lem}

\begin{cor}\label{cor beta} \textit{Let $\tau\in(0,1)$ be given. Assume conditions A1 - A3. Then, as $n\rightarrow \infty$,
	$$P[\hat{\beta}_{2.1}(\tau)\hat{\beta}_{1.2}(\tau)<0] \rightarrow 0.$$}
\end{cor}

From Lemma \ref{lem beta} and Corollary \ref{cor beta}, applying the multivariate $\delta$-method, we get the limiting normality of $\hat{\rho}_\tau$.

\begin{thm} \label{thm rhoq} \textit{Assume conditions A1 - A3. As $n \rightarrow \infty$, given $\tau \in (0,1)$, we have
\begin{equation}\nonumber
\begin{aligned}
\sqrt{n}(\hat{\rho}_\tau-\rho_\tau) \goesd N(0, V(\rho_\tau)).
\end{aligned}\end{equation}}\end{thm}
Theorem \ref{thm rhoq} is useful in constructing the statistical inference for the quantile correlation $\rho_\tau$ such as statistical significance of $\hat{\rho}_\tau$ and hypothesis tests. One can compute the p-value of the sample $\tau$-quantile correlation coefficient $\hat{\rho}_\tau$ by $p = 2\Phi(-|\hat{\rho}_\tau|/se(\hat{\rho}_\tau)),$ where $\Phi(\cdot)$ is the distribution function of the standard normal distribution. A valid $(1-\alpha)$-confidence interval of $\rho_\tau$ will be 
\begin{equation} \label{CI}
(\hat{\rho}_\tau-z_{\alpha/2}se(\hat{\rho}_\tau),~\hat{\rho}_\tau+z_{\alpha/2}se(\hat{\rho}_\tau)),~~\alpha \in (0,1),
\end{equation}
where $z_\alpha$ is the $\alpha$-quantile of the standard normal distribution. 
Furthermore, we can develop tests for tail dependence and asymmetry.

With $(\tau_1, \tau_2)=(\tau,0.5)$ or $(\tau,1-\tau)$, the tests of the tail dependence $\rho_\tau^D$ and tail correlation asymmetry $\rho_\tau^A$ are constructed from the asymptotic distribution of $\hat{\rho}_{\tau_1} - \hat{\rho}_{\tau_2}$ which is derived easily by applying the multivariate $\delta$-method to the limiting distribution of $\hat{\Theta}(\tau_1) - \hat{\Theta}(\tau_2)$ as given in Lemma \ref{lem asym two tau}. 

\begin{lem}\label{lem asym two tau} \textit{Let $\tau_1,\tau_2\in(0,1)$ be given. Assume conditions A1- A3 for $\tau=\tau_1,~\tau_2$. Then as $n\rightarrow\infty$,
	$$\sqrt{n}\begin{pmatrix} \hat{\Theta}(\tau_1) - \Theta(\tau_1) \\ 
	\hat{\Theta}(\tau_2) - \Theta(\tau_2) \end{pmatrix} \goesd {\color{black}N_8} (0, V({M}, H_{\tau_1\tau_1}, H_{\tau_1\tau_2}, H_{\tau_2\tau_1}, H_{\tau_2\tau_2})).$$}
\end{lem}

\begin{thm}\label{thm asym two tau} \textit{Under the same conditions for Lemma \ref{lem asym two tau}, as $n\rightarrow \infty$, we have
\begin{equation}\nonumber
\begin{aligned}
 \sqrt{n}( (\hat{\rho}_{\tau_1}-\rho_{\tau_1}) -(\hat{\rho}_{\tau_2}-\rho_{\tau_2})) \goesd N(0, V(\rho_{\tau_1}-\rho_{\tau_2})).
\end{aligned}
\end{equation}}
\end{thm}
From Theorem \ref{thm asym two tau}, given $\tau\in(0,0.5)$, valid tests for the tail {\color{black}independent} correlation	$H_0:\rho_\tau^D=0$ and for the tail correlation symmetry $H_0:\rho_\tau^A=0$ are $$t_\tau^D =\frac{\hat{\rho}_\tau^D}{se(\hat{\rho}_\tau^D)}~\text{ and }~t_\tau^A =\frac{\hat{\rho}_\tau^A}{se(\hat{\rho}_\tau^A)},$$
respectively, where $se(\hat{\rho}_\tau^D)$ and $se(\hat{\rho}_\tau^A)$ are given in (\ref{TD se}).   Under the null hypotheses, both of the two tests converge to the standard normal distribution as stated in the following corollaries. 
{\color{black}
\begin{cor} \label{test TD}
	\textit{Let $\tau\in (0,0.5)$ be given. Assume conditions A1 - A3 hold for $\tau$, $0.5$. Assume $\hat{M}$ with $\hat{f}_{Y_i\cdot X_i}(\hat{\theta}_{2.1}'(t)X_i^*),$ $\hat{f}_{X_i\cdot Y_i}(\hat{\theta}_{1.2}'(t)Y_i^*)$, $i=1,\cdots,n$, $t=\tau, 0.5$, are consistent for $M$. Then, under $\rho_{\tau}^D=0$, as $n\rightarrow\infty$, 
	$$t_\tau^D\goesd N(0,1).$$}
\end{cor}}
\begin{cor}\label{test TA} 
	\textit{Let $\tau \in (0,0.5)$ be given. Assume conditions A1 - A3 hold for $\tau$, $1-\tau$. Assume $\hat{M}$ with $\hat{f}_{Y_i\cdot X_i}(\hat{\theta}_{2.1}'(t)X_i^*)$, $\hat{f}_{X_i\cdot Y_i}(\hat{\theta}_{1.2}'(t)Y_i^*)$, $t=\tau,1-\tau$, are consistent for $M$. Then, under $\rho_\tau^A=0$, as $n\rightarrow \infty$,
	$$t_\tau^A \goesd N(0,1).$$}
\end{cor}
The conditional density estimators $\hat{f}^{BH}_{Y_i\cdot X_i}$, $\hat{f}^{BH}_{X_i \cdot Y_i}$ of Hyndman et al. (1996)  are consistent {\color{black}under some mild conditions on $a_{Y\cdot X}$, $b_{Y\cdot X}$, $a_{X\cdot Y}$, $b_{X\cdot Y}$ and $\hat{f}_{Y_i\cdot X_i}^{HK}$ and $\hat{f}_{X_i\cdot Y_i}^{HK}$ of Hendricks and Koenker (1991) are consistent under some mild conditions on $h_n$ and the additional linearity condition (\ref{quant eq}) on $Q_\tau^{Y_i}(X_i)$ and $Q_\tau^{X_i}(Y_i)$.} For more discussion on the consistency, see Hyndman et al. (1991) and Koenker (2005, p.77)

\section{Simulation} \label{sec simul}
We investigate finite sample validity of the asymptotic distribution of the sample quantile correlation $\hat{\rho}_\tau$ in Theorem \ref{thm rhoq} and finite sample performances of the proposed tests $t_\tau^D$ and $t_\tau^A$. The first one is checked by finite sample coverages of the $90\%$ confidence interval (CI) $[\hat{\rho}_\tau -1.645 se(\hat{\rho}_\tau), \hat{\rho}_\tau +1.645se(\hat{\rho}_\tau)]$ of the quantile correlation $\rho_\tau$. According to Theorem \ref{thm rhoq}, this CI is asymptotically valid. The second one is verified by size and power performances of the tests. 

We consider 
\begin{equation}\nonumber
\begin{aligned}
D_N & \text{: bivariate normal distribution},~~(X_i,Y_i)\sim \text{iid}~ N_2(0,(\begin{smallmatrix} 1 & \rho \\ \rho &1\end{smallmatrix})),~~\rho=0.5,\\
D_T & \text{: bivariate t distribution with 10 degree of freedom},~ t_{10},~~(X_i,Y_i)\sim \text{iid}~ t_{2}(0,(\begin{smallmatrix} 1 &\rho \\ \rho&1 \end{smallmatrix})),~~\rho=0.5,\\
D_R & \text{: iid from the rocket-type distribution in Example 2.1},\\
D_C & \text{: iid from the distribution of $(X_i,Y_i)$ in the cubic relation $Y=X^3+e$ in Example 2.2},\\
D_G & \text{: GARCH (1,1) model for $(X_{i},Y_{i})'=(\sigma_{Xi}e_{Xi}, \sigma_{Yi}e_{Yi})'$  with $ (e_{Xi},e_{Yi})'\sim \text{iid}~ N_2(0,(\begin{smallmatrix} 1 & \rho \\ \rho &1\end{smallmatrix})),~\rho=0.5$},\\
\end{aligned}
\end{equation}
from which  the data $\{(X_i,Y_i),i=1,\cdots,n\}$  are generated, $n=100,~500,~2500$. In $D_G$, GARCH (1,1) model is given as $\sigma_{Xi}^2 = 0.001+0.1X_{i-1}^2+0.85\sigma_{X,i-1}^2$ and  $\sigma_{Yi}^2 = 0.001+0.1Y_{i-1}^2+0.85\sigma_{Y,i-1}^2$. The normal and t random variables are generated by ``rmvnorm" and ``rmvt" in the R package.

Note that $(X_i,Y_i)$ from $D_N$, $D_T$, $D_R$, $D_C$ is a sequence of iid random variable satisfying condition A1(i) and that $(X_i,Y_i)$ from $D_G$ is a sequence of non-iid martingale difference satisfying condition A1(ii).
According to Theorem \ref{thm norm}, $D_N$ has constant quantile correlation $\rho_\tau=\rho=0.5$ for all $\tau \in (0,1)$ and almost so has $D_T$ with $\rho_\tau=0.500$ for $\tau=(0.1,0.5,0.9)$. Other DGPs have tail-dependent $\rho_\tau$: $D_R$ has asymmetric tail dependent quantile correlation; $D_C$ has symmetric tail dependent quantile correlation; $D_G$ has $\rho=0.483$ and $\rho_\tau = (0.487,0.471, 0.487)$ for $\tau=(0.1,0.5,0.9)$. For each of the five distributions, $M=10000$ independent $\hat{\rho}_{\tau i},~i=1,\cdots,M$, and the corresponding CIs of $\rho_\tau$ are constructed, $\tau=0.1, 0.5, 0.9$. 

In order to construct the CIs and test statistics, we need estimators $\hat{f}_{Y_i\cdot X_i}(\theta_{2.1}'X_i^*)$ and $\hat{f}_{X_i\cdot Y_i}(\theta_{1.2}'Y_i^*)$ of the conditional density values for which we consider the two strategies of (\ref{cond dist2}) by Hyndman et al. (1996) and (\ref{cond dist}) by Hendricks and Koenker (1991).
The conditional kernel density of Hyndman et al. (1996) is (\ref{cond dist2}) with the Gaussian kernel, $$K(u)=\phi(u)=exp(-u^2/2)/\sqrt{2\pi},$$ and  bandwidths
{\small \begin{equation}\label{ab}
a_{Y\cdot X}=\{\frac{16kR^2(K)\hat{p}_Y^5(288\pi^9\hat{\sigma}_{XX}^{58}\lambda^2(k))^{1/8}}{n\hat{d}_{Y\cdot X}^{5/2}\hat{v}^{3/4}_{Y\cdot X}(k)[\hat{v}^{1/2}_{Y\cdot X}(k)+\hat{d}_{Y\cdot X}(18\pi\hat{\sigma}_{XX}^{10}\lambda^2(k))^{1/4}]}\}^{1/6},~~b_{Y\cdot X}=\{\frac{\hat{d}_{Y\cdot X}^2\hat{v}_{Y\cdot X}(k)}{3\sqrt{2\pi}\hat{\sigma}_{XX}^5\lambda(k)}\}^{1/4}a_{Y\cdot X},
 \end{equation}}
where $$\lambda(k) = \Phi(k)-\Phi(-k),~~k=3,~~ R(K)=\int_{-\infty}^\infty K^2(w)dw,~~\hat{d}_{Y\cdot X}=\hat{\rho}\frac{\hat{\sigma}_{YY}}{\hat{\sigma}_{XX}},~~\hat{p}_Y=\sqrt{(1-\hat{\rho}^2)\hat{\sigma}_{YY}^2},$$$$\hat{v}_{Y\cdot X}(k) = \sqrt{2\pi}\hat{\sigma}_{XX}^3(3\hat{d}_{Y\cdot X}^2\hat{\sigma}_{XX}^2+8\hat{p}_Y^2)\lambda(k)-16k\hat{\sigma}_{XX}^2\hat{p}_Y^2e^{-k^2/2},$$ $\hat{\sigma}_{XX}^2$ and $\hat{\sigma}_{YY}^2$ are the sample variances of $X$ and $Y$, respectively. The other bandwidth parameters $a_{X\cdot Y}$, $b_{X\cdot Y}$ are computed by interchanging $X$, $Y$ in (\ref{ab}). According to Bashtannyk and Hyndman (2001), the bandwidths are estimated values of the optimal bandwidths for bivariate normal distributions.

The estimated conditional density values proposed by Hendricks and Koenker (1991) are (\ref{cond dist}) with $\epsilon=0.001$ and bandwidth
\begin{equation}\label{h}
h_n=n^{-1/5}(\frac{4.5\phi^4(\Phi^{-1}(\tau))}{(2\Phi^{-1}(\tau)^2+1)^2})^{1/5},
\end{equation}
where $\phi(\cdot)$ and $\Phi(\cdot)$ are pdf and cdf of the standard normal distribution, respectively. According to Bofinger (1975) and Koenker (2005, p.115), $h_n$ is optimal under normality of $(X,Y)$.

Empirical coverage probability of $90\%$ CI of the quantile correlation coefficient $\rho_\tau$ is 
\begin{equation} \label{coverage}
coverage = \frac{1}{M}\sum_{i=1}^M I[\hat{\rho}_{\tau i} -1.645 se(\hat{\rho}_{\tau i})<\rho_\tau< \hat{\rho}_{\tau i}+1.645se(\hat{\rho}_{\tau i})],
\end{equation}
where $I(A)$ is the indicator function of an event $A$. 
The true values $\rho_\tau$ for $D_R$ and $D_C$ are given in Table 1 and Table 2, those for $D_N$ and $D_T$ are $\rho_\tau=\rho=0.5$.

Table 4 shows the coverages and lengths of  the 90\% CIs.
Consider first the results with the conditional kernel density $\hat{f}^{BH}$ in (\ref{cond dist2}) of Hyndman et al. (1996).
Except for $D_C$ with $\tau=0.1,0.9$, for all five DGPs, the proposed CI of $\rho_\tau$ has coverages not much deviated from the given value 90\% even though there are some under-coverages for $\tau=0.1,0.9$, which generally improve to the nominal coverage 90\% as $n$ increases from 100 to 500 and next to 2500. The under-coverage problems is caused by inefficient estimate of conditional  density values due to the small number of sample in tails especially for $\tau=0.1,0.9$. 
The degree of under-coverage is similar to that the CI of the coefficient $\beta_{2.1}$ of the quantile regression $Q_\tau^Y(X)=\alpha_{2.1}+\beta_{2.1}X$, reported in the Monte-Carlo studies of Koenker (2005, Section 3.10) and Kocherginskty et al. (2005, Section 4).

Consider next the results with the conditional kernel density values $\hat{f}^{HK}$  in (\ref{cond dist}) of Hendricks and Koenker (1991). Except for $D_C$, for all $n$, $\tau$ considered here, coverage of the proposed CI of $\rho_\tau$ is generally acceptable and is better than the corresponding coverage based on the other conditional density estimator $\hat{f}^{BH}$. 
Regarding average lengths of CIs, that based on $\hat{f}^{BH}$ tends to be shorter in tails and longer in center than that based on $\hat{f}^{HK}$, which however dues to smaller coverage of them in tails and larger coverage in center. Therefore, none of $\hat{f}^{BH}$ and $\hat{f}^{HK}$ {\color{black}are} better than the other one in average length of CI.

{\scriptsize
	\begin{center}
		
		{\bf Table 4.}  {\footnotesize Empirical coverages (\%) and lengths of the 90\% confidence intervals of {\color{black}quantile} correlation coefficients}
		
		{\begin{tabular}{c|l|ccc|ccc|ccc|ccc}
				\hline
				&&\multicolumn{6}{c}{Coverage (\%)} \vline& \multicolumn{6}{c}{Length}\\
				&&\multicolumn{3}{c}{$\hat{f}^{BH}$} \vline& \multicolumn{3}{c}{$\hat{f}^{HK}$}\vline&\multicolumn{3}{c}{$\hat{f}^{BH}$} \vline& \multicolumn{3}{c}{$\hat{f}^{HK}$} \\
				$n$ &  \multicolumn{1}{c}{DGP} \vline & $\rho_{0.1}$ & $\rho_{0.5}$ & $\rho_{0.9}$ & $\rho_{0.1}$ & $\rho_{0.5}$ & $\rho_{0.9}$ & $\rho_{0.1}$ & $\rho_{0.5}$ & $\rho_{0.9}$ & $\rho_{0.1}$ & $\rho_{0.5}$ & $\rho_{0.9}$\\
				\hline
				100 &$D_N$, Normal  &83.8 &93.4 &87.5 &88.5 &91.3 &91.2 &0.31 &0.34 &0.35 &0.38 &0.32 &0.42\\
				&$D_T$, $t_{10}$    &83.5 &93.1 &86.9 &91.0 &87.8 &92.4 &0.35 &0.35 &0.39 &0.47 &0.29 &0.52\\
				&$D_R$, Example 2.1 &77.5 &93.8 &88.5 &85.2 &90.6 &91.0 &0.33 &0.35 &0.35 &0.40 &0.32 &0.41\\
				&$D_C$, Example 2.2 &66.8 &91.7 &70.5 &82.0 &72.3 &85.3 &0.25 &0.27 &0.27 &0.44 &0.16 &0.48\\
				&$D_G$: GARCH(1,1)  &83.8 &92.0 &87.2 &89.4 &88.5 &91.6 &0.32 &0.34 &0.36 &0.40 &0.31 &0.44\\
				\hline
				500 &$D_N$: Normal  &87.0 &93.5 &88.0 &90.1 &91.6 &91.1 &0.15 &0.15 &0.15 &0.16 &0.14 &0.17\\
				&$D_T$: $t_{10}$    &86.6 &92.5 &88.0 &91.5 &87.2 &92.4 &0.16 &0.15 &0.17 &0.19 &0.13 &0.20\\
				&$D_R$: Example 2.1 &85.5 &93.1 &87.7 &88.8 &90.4 &90.0 &0.17 &0.15 &0.15 &0.19 &0.14 &0.16\\
				&$D_C$: Example 2.2 &82.2 &91.8 &82.9 &81.0 &66.9 &81.9 &0.16 &0.11 &0.16 &0.17 &0.06 &0.17\\
				&$D_G$: GARCH(1,1)  &86.3 &91.4 &87.0 &90.1 &87.9 &90.5 &0.15 &0.14 &0.15 &0.17 &0.13 &0.17\\
				\hline
				2500 &$D_N$: Normal  &89.1 &92.6 &88.3 &91.0 &91.2 &90.6 &0.07 &0.06 &0.07 &0.07 &0.06 &0.07\\
				&$D_T$: $t_{10}$    &88.7  &91.0  &88.5  &92.5  &85.8  &92.4  &0.08  &0.06  &0.08  &0.09  &0.06  &0.09\\
				&$D_R$: Example 2.1 &88.7 &91.7 &89.1 &90.9 &89.8 &90.5 &0.08 &0.06 &0.07 &0.09 &0.06 &0.07\\
				&$D_C$: Example 2.2 &86.8 &90.6 &86.6 &76.3 &64.9 &76.6 &0.08 &0.05 &0.08 &0.06 &0.03 &0.06\\
				&$D_G$: GARCH(1,1)  &87.8 &89.6 &87.5 &90.6 &85.9 &90.4 &0.07 &0.06 &0.07 &0.08 &0.06 &0.08\\
				\hline
		\end{tabular}}
		
		{\scriptsize Note: $\hat{f}^{BH}=(\hat{f}^{BH}_{Y\cdot X},\hat{f}^{BH}_{X\cdot Y})$ is conditional kernel density of Hyndman et al. (1996).  $\hat{f}^{HK}=(\hat{f}^{HK}_{Y\cdot X},\hat{f}^{HK}_{X\cdot Y})$ is conditional kernel density of Hendricks and Koenker (1991). }
	\end{center}
}

Finite sample size and power studies of the proposed tests $t_\tau^D$ and $t_\tau^A$, $\tau=0.1,$ are made. Rejection rates of the tests out of $M=10000$ independent replications are reported in Table 5. For $D_N$ and $D_T$, the rejection rates of $\rho_\tau^D$ and $\rho_\tau^A$ are both sizes; for $D_R$,  those for $t_\tau^A$ and $t_\tau^D$ are powers; and for $D_C$ {\color{black}and $D_G$}, that for $t_\tau^A$ is size and that for $t_\tau^D$ is power.

Except for $t_{0.1}^A$ for $D_C$, the table shows reasonable sizes, which improves to the given level 5\% as in $n$ increases from 100 to 500 and next to 2500. For $D_C$, size of $t_{0.1}^A$ based on $\hat{f}^{BH}$ is poor for $n=100$, which however improves rapidly as $n$ increases from 100 to 500 and on. This bad performance of $t_{0.1}^A$ for $n=100$ is a consequence of inefficiency of $\hat{f}^{BH}$. Since $\hat{f}^{BH}$ is consistent, the bad coverage of $t_{0.1}^A$ for $D_C$ disappears as $n$ increases to 500. On the other hand, for $D_C$, size of $t_{0.1}^A$ based on $\hat{f}^{HK}$ is not so bad for $n=100,500$, but is bad for $n=2500$. The bad performance for $n=2500$ is a consequence of nonlinearity of the quantile functions of $D_C$. 

The table shows the proposed tests have powers which increases as $n$ increases from 100 to 2500. The power values are not large even for $n=2500$. This implies that we need a large sample in order to detect tail dependency or tail asymmetry if any. 

\newpage
{\scriptsize
	\begin{center}
		
		{\bf Table 5.}  {\footnotesize Sizes (\%) and powers (\%) of level 5\% tests}
		\scalebox{1}{
			\begin{tabular}{c|l|ccc|ccc}
				\hline
				&&\multicolumn{3}{c}{$t_{0.1}^D$}& \multicolumn{3}{c}{$t_{0.1}^A$}\\
				n & \multicolumn{1}{c}{DGP} \vline& size/power & $\hat{f}^{BH}$ &$\hat{f}^{HK}$ & size/power & $\hat{f}^{BH}$ &$\hat{f}^{HK}$ \\ 
				\hline
				100 &$D_N$: Normal  &size  &3.9 &3.3 &size &6.4 &3.6\\
				&$D_T$: $t_{10}$    &size  &4.5 &3.1 &size &7.6 &2.6\\
				&$D_R$: Example 2.1 &power &5.6 &4.5 &power &8.8 &4.9\\
				&$D_C$: Example 2.2 &power &13.2 &10.6 &size &29.5 &8.2\\
				&$D_G$: GARCH(1,1)  &power &4.0 &3.6 &size &6.6 &3.3\\
				\hline
				500 &$D_N$: Normal  &size  &4.8 &4.2 &size &6.2 &4.3\\
				&$D_T$: $t_{10}$    &size  &5.0 &3.9 &size &6.2 &3.1\\
				&$D_R$: Example 2.1 &power &7.7 &6.4 &power &8.8 &5.8\\
				&$D_C$: Example 2.2 &power &10.7 &15.4 &size &10.5 &9.2\\
				&$D_G$: GARCH(1,1)  &power &6.4 &5.7 &size &6.4 &4.2\\
				\hline
		       2500 &$D_N$: Normal  &size  &4.8 &4.5 &size &5.5 &4.4\\
				&$D_T$: $t_{10}$    &size  &5.3 &4.2 &size &5.4 &3.3\\
				&$D_R$: Example 2.1 &power &18.8 &17.0 &power &17.7 &14.9\\
				&$D_C$: Example 2.2 &power &17.1 &31.1 &size &7.6 &14.5\\
				&$D_G$: GARCH(1,1)  &power &12.4 &11.8 &size &5.0 &3.8\\
				\hline
		\end{tabular}}
	
	{\scriptsize Note: $\hat{f}^{BH}=(\hat{f}^{BH}_{Y\cdot X},\hat{f}^{BH}_{X\cdot Y})$ is conditional kernel density of Hyndman et al. (1996).  $\hat{f}^{HK}=(\hat{f}^{HK}_{Y\cdot X},\hat{f}^{HK}_{X\cdot Y})$ is conditional kernel density of Hendricks and Koenker (1991). }
	\end{center}
}
We discuss relative performance of $\hat{f}^{BH}$ and $\hat{f}^{HK}$. Consider first $n=500,2500$. The density estimator $\hat{f}^{BH}$ gives always-acceptable coverages for the CI and sizes for the tests $t_{0.1}^{D}$, $t_{0.1}^{A}$, while the other one $\hat{f}^{HK}$ gives poor coverages for CI and sizes for $t_{0.1}^A$ for $D_C$.
Consider next $n=100$.
We observe stable coverages of the proposed CIs and sizes of the proposed tests $t_\tau^D$ and $t_\tau^A$ for the last four DGPs, $D_T$, $D_R$, $D_C$, $D_G$, except for $t_\tau^A$ for $D_C$ even though we have used the conditional density value estimators $\hat{f}^{HK}$. 
For samples of not small size, we recommend the always-consistent $\hat{f}^{BH}$, but, for samples of small size, one may better use $\hat{f}^{HK}$.

We summarize the results of this Monte-Carlo simulation. First, the confidence interval of ${\rho}_\tau$ has reasonable finite sample coverages and the proposed tests $t_\tau^D$, $t_\tau^A$ have generally  acceptable sizes and power for the bivariate distributions with tail dependent correlation considered here. This fact confirms both finite sample validity of the asymptotic theory and usefulness of the proposed methods based on $\hat{\rho}_\tau$. 
Second, among the two conditional density value estimators considered here, for samples of not small size, those $\hat{f}^{BH}$ by Hyndman et al. (1996) provide the confidence intervals and tests with better finite sample coverage and sizes than the other one $\hat{f}^{HK}$ by Hendricks and Koenker (1991), while, for samples of small size, $\hat{f}^{HK}$ is better than $\hat{f}^{BH}$.

\section{Real data set analysis}\label{sec data}

The proposed quantile correlation methods are applied a couple of field data sets: birth weight data set and stock return data set. The data analysis illustrates well the tail-dependent sensitivity of the conditional quantile of one variable with respect to change of the other variable.

\subsection{Birth weight data}
We first analyze the birth weight data set considered by Abreveya (2001) and Koenker (2005) for identifying impact factors on the birth weight. The data set is the natality data published by the US national center for health statistics in June 2017. We investigate the relationship between mother's weight $(X)$ gained during pregnancy and birth weight $(Y)$. We have $n=50000$. The sample may be regarded to be iid satisfying condition A1(i). Figure \ref{Fig birth} (a) displays a scatter plot of birth weight and mother's weight gain. 
The figure shows an overall positive correlation.

We compute sample quantile correlation $\hat{\rho}_\tau$. We also compute the sample quantile correlations $\hat{\rho}_{I\tau}^{X,Y},~ \hat{\rho}_{I\tau}^{Y,X}$ of Li et al. (2015), which are the Pearson correlations of $X^I=I(X>Q_\tau^X)$ and $Y$ and of $Y^I=I(Y>Q_\tau^Y)$ and $X$, respectively, where $Q_\tau^X$ and $Q_\tau^Y$ are the $\tau$-quantiles of $X$ and $Y$. 

\begin{figure}[h]                                                                   %
	\centering                                                                          %
	\includegraphics[width=1\textwidth]{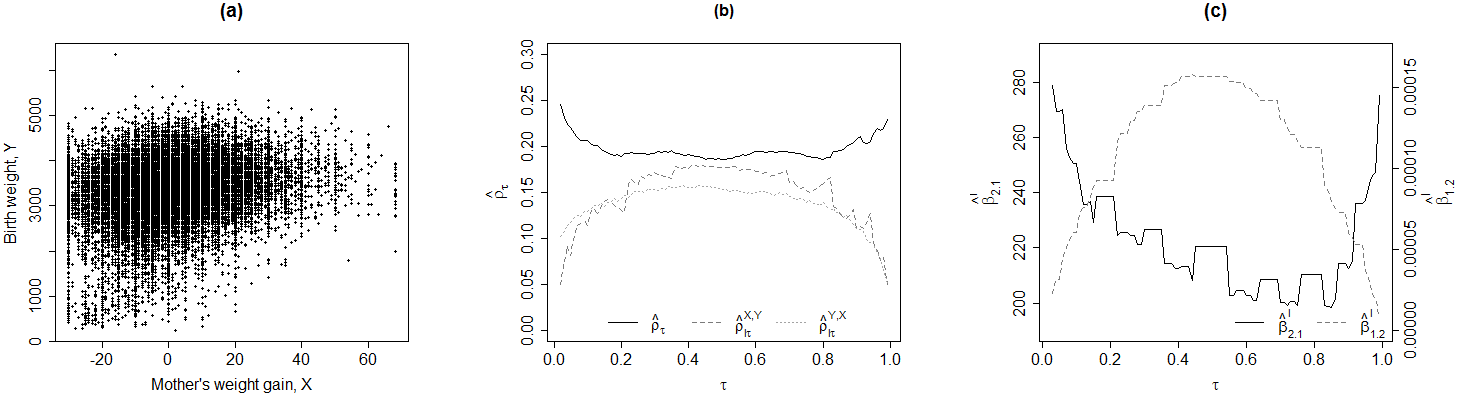}
	\caption{(a) scatter plot of X=mother's weight gain and Y=birth weight; (b) sample quantile correlation $\hat{\rho}_\tau$, $\hat{\rho}_{I\tau}^{X,Y}$ and $\hat{\rho}_{I\tau}^{Y,X}$; (c) $\hat{\beta}_{1.2}^I(\tau)$ and $\hat{\beta}_{2.1}^I(\tau)$ in $\hat{\rho}_{I\tau}^{X,Y}=\widehat{corr}(I(X>\hat{Q}_\tau^X),Y)=\sqrt{\hat{\beta}_{1.2}^I(\tau)\hat{\beta}_{2.1}^I(\tau)}$ }\label{Fig birth}                         %
\end{figure}

Figure \ref{Fig birth} (b) shows the sample quantile correlation coefficient $\hat{\rho}_\tau$, $\tau=0.01, 0.02, \cdots, 0.99$, which is superimposed on $\hat{\rho}_{I\tau} = (\hat{\rho}_{I\tau}^{X,Y},  \hat{\rho}_{I\tau}^{Y,X})$.  We note a tail-dependent convex feature that $\hat{\rho}_\tau$, $\hat{\rho}_{1-\tau}$ in tails of $\tau \le 0.1$  are greater than $\hat{\rho}_{0.5}$. This means that conditional tail $\tau-$quantile, $\tau \le 0.1$, $\tau \ge 0.9$, of one variable varies more sensitively with change of the other variable than the conditional median of the variable. The other quantile correlations $\hat{\rho}_{I\tau}^{X,Y}$ and $\hat{\rho}_{I\tau}^{Y,X}$ tell a quite different story having concave shapes. Figure 2 (b) shows that $\hat{\rho}_{I\tau}^{X,Y}$ and $\hat{\rho}_{I\tau}^{Y,X}$ are closer to 0 at tails than at center. This indicates that the indicator $X^I$ and $Y$ or $X$ and the indicator $Y^I$ have stronger association at center.

Concavity of $\hat{\rho}_{I\tau}^{X,Y}=\sqrt{\hat{\beta}_{2.1}^I(\tau)\hat{\beta}_{1.2}^I(\tau)}$ is a compound result of convexity of  $\hat{\beta}_{2.1}^I(\tau)=\hat{E}[Y|X > \hat{Q}_\tau^X]-\hat{E}[Y|X\le\hat{Q}_\tau^X]$ and concavity of $\hat{\beta}_{1.2}^I(\tau)$, the slope of $Y$ in the regression of $I(X> \hat{Q}_\tau^X)$ on $Y$, whose plots are given in Figure \ref{Fig birth} (c). Note that $\hat{\beta}_{2.1}^I(\tau)$ is heterogeneity in conditional means, $\hat{\beta}_{1.2}^I(\tau)$ is sensitivity of the conditional probability $P(X> Q_\tau^X|Y)$ to change of $Y$, and  they are reversely shaped. Therefore, it is hard to get a simple interpretation from $\hat{\rho}^{X,Y}_{I\tau}$ and so is from $\hat{\rho}_{I\tau}^{Y,X}$.

{\scriptsize
	\begin{center}
		{\bf Table 6.}  {\footnotesize Estimated $\hat{\rho}_\tau^D$ and $\hat{\rho}_\tau^A$, and t-tests statistics $t_\tau^D$ and $t_\tau^A$}
		\begin{tabular}{c|ccc|ccc|ccc}
			\hline
			&\multicolumn{3}{c }{$\tau=0.05$} \vline&\multicolumn{3}{c }{$\tau=0.10$} \vline&\multicolumn{3}{c }{$\tau=0.25$} \\
			&estimate & t-stat &p-value &estimate & t-stat &p-value &estimate & t-stat &p-value\\
			\hline
			$\rho_\tau^D$ &0.034 &4.733 &$<0.001$ &0.020 &4.126 &$<0.001$ &0.006 &1.687 &$0.092$ \\ 
			$\rho_\tau^A$ &0.006 &0.684 &0.494 &-0.001 &-0.179 &0.858 &0.003 &0.660 &0.509\\
			\hline
		\end{tabular}
	\end{center}
}

\vspace{2mm}

Table 6 provides estimated measures $\hat{\rho}_\tau^D$ and $\hat{\rho}_\tau^A$, their t-test statistics, and their p-values of the test $t_\tau^D$ and $t_\tau^A$, $\tau=0.05,0.10,0.25$ in which the conditional density estimator $\hat{f}^{BH}$ is used according to the recommendation of Section \ref{sec simul} for $n$ not small. 
The differences between median quantile correlation $\hat{\rho}_{0.5}$ and lower and upper quantile correlation $\hat{\rho}_\tau$ shown in Figure {\ref{Fig birth}} (b) are significant with p-value of $t_\tau^D<0.05,~\tau=0.05, 0.1$. Note that $\hat{\rho}_\tau$ is symmetric in that lower quantile $\hat{\rho}_{\tau}$ and upper quantile $\hat{\rho}_{1-\tau}$ are not significantly different from each other having p-values of $t_\tau^A$ larger than 0.05, for $\tau=0.05, 0.10, 0.25$.

\subsection{Stock return data}
We next analyze asymmetric tail dependent relations of pairs of log returns of two stock price indices for the period of 01/03/2000 - 11/30/2017: the US S\&P 500 index and the French CAC 40 index. The stock price data sets are obtained from the Oxford-Man realized library (http://realized.oxford-man.ox.ac.uk). We have $n= 4072$. The sample may be regarded from a martingale difference satisfying condition A1(ii). Confidence intervals of $\rho_\tau$ for $\tau$ closer to 0 or 1 are wider than those of $\rho_\tau$ for $\tau$ close to 0.5. This implies that $\hat{\rho}_\tau$ for $\tau$ closer to 0 or 1 has larger standard error than $\hat{\rho}_\tau$ for $\tau$ closer to 0.5.

\begin{figure}[h]                                                                   %
	\centering                                                                          %
	\includegraphics[width=0.4\textwidth]{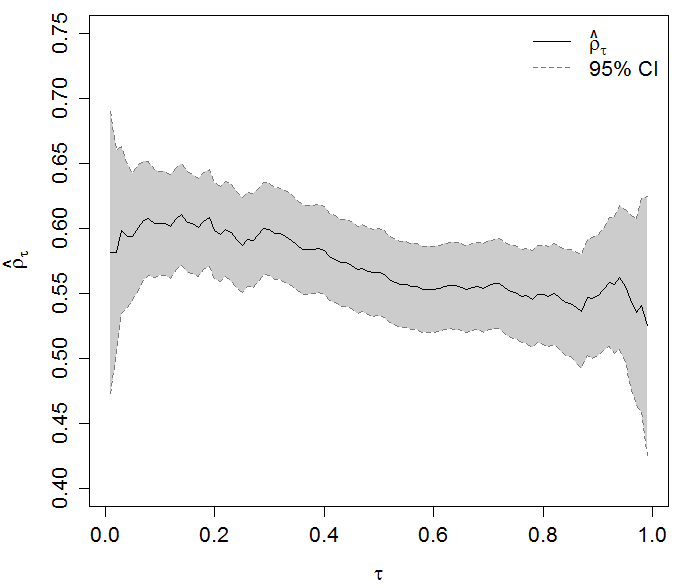}
	\caption{Sample quantile correlations $\hat{\rho}_\tau$ and 95\% confidence interval (CI) for $\hat{\rho}_\tau$} \label{Fig quant corr}                         %
\end{figure}  

Figure \ref{Fig quant corr} reports the sample quantile correlation coefficient $\hat{\rho}_\tau$ and 95\% confidence interval for $\hat{\rho}_\tau$  constructed from (\ref{CI}), $\tau=0.01, 0.02,$ $\cdots, 0.99$ in which $\hat{f}^{BH}$ is used according to the recommendation of it for $n$ not small. 
We see strongly tail-dependent $\hat{\rho}_\tau$: roughly, $\hat{\rho}_\tau>\hat{\rho}_{0.5}\backsimeq \hat{\rho}>\hat{\rho}_{1-\tau},~\tau<0.5$. This implies that the left conditional quantile of one stock return is more sensitive to change of the other return than its conditional median and the right conditional quantile is less sensitive than its conditional median. For example, the 5\% conditional VaR (value-at-risk) of one stock return is more sensitive to change of the other stock return than its conditional median. The left 5\% conditional VaR of one stock return is more sensitive to the other stock return than the right 5\% conditional VaR.

{\scriptsize
	\begin{center}
		{\bf Table 7.}  {Estimated $\hat{\rho}_\tau^D$ and $\hat{\rho}_\tau^A$, and t-tests statistics $t_\tau^D$ and $t_\tau^A$}
		
		\begin{tabular}{c|ccc|ccc|ccc}
			\hline
			&\multicolumn{3}{c }{$\tau=0.05$} \vline&\multicolumn{3}{c }{$\tau=0.10$} \vline&\multicolumn{3}{c }{$\tau=0.25$} \\
			&estimate & t-stat &p-value &estimate & t-stat &p-value &estimate & t-stat &p-value\\
			\hline
			$\hat{\rho}_\tau^D$ &0.027 &1.263 &0.207 &0.037 &2.144 &0.032 &0.021 &1.807 &0.071\\ 
			$\hat{\rho}_\tau^A$ &0.038 &1.248 &0.212 &0.055 &2.176 &0.030 &0.036 &2.261 &0.024\\
			\hline
		\end{tabular}
	\end{center}
}

\vspace{2mm}

This asymmetric feature of $\hat{\rho}_\tau$ is demonstrated by measures $\hat{\rho}_\tau^D$ and $\hat{\rho}_\tau^A$ and tests $t_\tau^D$ and $t_\tau^A$ reported in  Table 7, $\tau=0.05,~0.10,~0.25$. We see positive values for all $\hat{\rho}_\tau^D$ and $\hat{\rho}_\tau^A$, $\tau=0.05, 0.10,0.25,$ indicating more sensitive for left conditional quantile of a return to change of the other return than its conditional median return and than the corresponding right conditional quantile  of the return, respectively. The p-values of $t_\tau^D$ and $t_\tau^A$ $<0.10$ show significant tail dependence and asymmetry in the quantile correlation at the $\tau=0.1,~0.25$ tails. Insignificance of $t_\tau^D$ and $t_\tau^A$ at $\tau=0.05$ may be a consequence of the reduced sample sizes for the deep left tail of $\tau=0.05$.

\section{Conclusion}
We have proposed a new correlation measure in which tail dependence is reflected, called quantile correlation coefficient. It is defined to be the geometric mean of two quantile regression slopes of X on Y and Y on X. The proposed correlation coefficient is shown to share the basic properties of the usual Pearson correlation coefficient: zero for independent pairs of random variables, $\pm1$ for perfectly linearly related pairs, scale-location equivariance, and being bounded by $1$ in absolute value for general class of random pairs. 
Tail-dependent association of $X$, $Y$, if any, is well captured by the proposed quantile correlation coefficient.
The new measure allows us to measure sensitivity of conditional quantile of one variable with respect to change the other variable.
The quantile correlation coefficient is easy to estimate and clear to  interpret. Based on the quantile correlation coefficient, we proposed measure of tail dependent correlation and of tail correlation asymmetry and their statistical tests. We established asymptotic normalities for the sample quantile correlation coefficient and for the proposed tests under null hypothesis. A Monte-Carlo study confirms the asymptotic normality of the quantile correlation and shows reasonable sizes and powers for the proposed tests.
Birth weight data set and stock return data set were analyzed by the proposed quantile correlation methods to have tail dependent correlations. The analysis reveals that degrees of sensitivity for lower, upper, median conditional quantiles of one variable to change in other variable are different.

\clearpage

\section*{Appendix - proofs}
In the proofs of the theorems in Section 2, we use the following property:
for any $a>0$ and $\tau \in[0,1]$, $l_\tau(ae)=al_\tau(e)$, $l_\tau(-ae)=al_{1-\tau}(e)$ and we use $\beta_{2.1}=\beta_{2.1}(\tau)$, $\beta_{1.2}=\beta_{1.2}(\tau)$ for simplicity of notation.

\noindent\textbf{Proof of Theorem \ref{thm welldef}.}
Assume $\beta_{2.1}\beta_{1.2}<0.$ Let $\beta_{2.1} > 0$ and $\beta_{1.2}<0$. 
Then     
\begin{equation} \nonumber
\begin{aligned}
E[l_\tau(Y-&X\beta_{2.1}-\alpha_{2.1})]E[l_\tau(X-Y\beta_{1.2}-\alpha_{1.2})] \\
= & E[l_{1-\tau}(X-\frac{1}{\beta_{2.1}}Y+\frac{\alpha_{2.1}}{\beta_{2.1}})]E[l_{1-\tau}(\frac{1}{\beta_{1.2}}X-Y-\frac{\alpha_{1.2}}{\beta_{1.2}})]\beta_{2.1}\beta_{1.2}<0,
\end{aligned}
\end{equation}
which is a contradiction because $E[l_\tau(e)]\ge 0$ for all $e$. For $\beta_{2.1}<0$ and $\beta_{1.2} > 0$, the same contradiction is derived. 
Hence, $\beta_{2.1}\beta_{1.2}\ge0$. $\Box$

\vspace{3mm}

\noindent\textbf{Proof of Theorem \ref{coef}.}
By linearity assumption,
$Q_Y(\tau|X) = \alpha_{2.1}^\tau+\beta_{2.1}^\tau X$ and $Q_X(\tau|Y) = \alpha_{1.2}^\tau+\beta_{1.2}^\tau Y$
for $(\alpha_{2.1}^\tau,\beta_{2.1}^\tau)$ and $ (\alpha_{1.2}^\tau,\beta_{1.2}^\tau)$ which minimize 
{\color{black}$E[L_\tau^{X,Y}(\alpha,\beta)|X]$} for any $X$ and {\color{black}$E[L_\tau^{Y,X}(\alpha,\beta)|Y]$} for any $Y$, respectively. 
Note that $(\alpha_{2.1}^\tau,\beta_{2.1}^\tau),$ and $(\alpha_{1.2}^\tau,\beta_{1.2}^\tau)$ do not depend on $X$ and $Y$. 
Therefore, $(\alpha_{2.1}^\tau,\beta_{2.1}^\tau)$ and $(\alpha_{1.2}^\tau,\beta_{1.2}^\tau)$ minimize {\color{black}$L_\tau^{X,Y}(\alpha,\beta)$ and $L_\tau^{Y,X}(\alpha,\beta)$,} respectively, and hence 
$(\alpha_{2.1}^\tau,\beta_{2.1}^\tau)=(\alpha_{2.1}(\tau),\beta_{2.1}(\tau))$, $(\alpha_{1.2}^\tau,\beta_{1.2}^\tau)= (\alpha_{1.2}(\tau),\beta_{1.2}(\tau)).~\Box$

\vspace{3mm}

\noindent\textbf{Proof of Theorem \ref{thm prop}.} \textit{(i)} By Theorem \ref{thm welldef}, we have $$\rho_\tau^{X,Y} = sign(\beta_{2.1})\sqrt{\beta_{2.1}\beta_{1.2}} = sign(\beta_{1.2})\sqrt{\beta_{1.2}\beta_{2.1}}=\rho_\tau^{Y,X}.$$
 
 \textit{(ii)} Applying Koenker (2005, Theorem 2.3), we have
 $$\rho_\tau^{aX+b, cY+d} = sign(\frac{c}{a}\beta_{2.1})\sqrt{\frac{c}{a}\beta_{2.1}\frac{a}{c}\beta_{1.2}} = sign(\beta_{2.1})\sqrt{\beta_{2.1}\beta_{1.2}}=\rho_\tau^{X,Y}.$$

 \textit{(iii)} Since $E[l_\tau(Y-\alpha-\beta X)] \ge 0$, for all $\alpha, \beta, \tau$, $(\alpha_{2.1},\beta_{2.1})=(\gamma,\delta)$ minimizes $E[l_\tau(Y-\alpha-\beta X)]$ to $E[l_\tau(Y-\alpha_{2.1}-\beta_{2.1}X)]=0$. Similarly, from {\color{black}$X=-\frac{\gamma}{\delta}+\frac{1}{\delta}Y$,} $(\alpha_{1.2},\beta_{1.2})=(-\frac{\gamma}{\delta}, \frac{1}{\delta})$ minimizes $E[l_\tau(X-\alpha-\beta Y)]$ to $E[l_\tau(X-\alpha_{1.2}-\beta_{1.2}Y)] =0$ and we get the desired result. $\Box$ 

 \textit{(iv)} From independence of $X$ and $Y$, we have $Q_Y(\tau|X) = Q_Y(\tau)$ and $Q_X(\tau|Y)=Q_X(\tau)$, which are the conditional $\tau$-quantiles of $Y$ and $X$, respectively. By Theorem \ref{coef}, $\beta_{2.1}(\tau)=\beta_{1.2}(\tau)=0$ and hence $\rho_\tau=0.~\Box$

\vspace{3mm}
\noindent\textbf{Proof of Theorem \ref{thm 1}.}
Assume $\beta_{2.1}\beta_{1.2}>1$. Let \textit{(i)} hold that $\beta_{1.2}<0$, $\beta_{2.1}<0$. Then \begin{equation} \nonumber
\begin{aligned} E[l_\tau(Y-&\alpha_{2.1}-\beta_{2.1}X)]E[l_\tau(X-\alpha_{1.2}-\beta_{1.2}Y)] \\
= &\beta_{2.1}\beta_{1.2}E[l_\tau(X+\frac{\alpha_{2.1}}{\beta_{2.1}}-\frac{1}{\beta_{2.1}}Y)]
E[l_\tau(Y+\frac{\alpha_{1.2}}{\beta_{1.2}}-\frac{1}{\beta_{1.2}}X)]\\
>&E[l_\tau(X+\frac{\alpha_{2.1}}{\beta_{2.1}}-\frac{1}{\beta_{2.1}}Y)]E[l_\tau(Y+\frac{\alpha_{1.2}}{\beta_{1.2}}-\frac{1}{\beta_{1.2}}X)]
\end{aligned}
\end{equation}
which is a contradiction because $(\alpha_{2.1},\beta_{2.1})=argmin_{(\alpha,\beta)}L_\tau^{X,Y}(\alpha,\beta)$ and $(\alpha_{1.2},\beta_{1.2})=argmin_{(\alpha, \beta)}L_\tau^{Y,X}(\alpha,\beta)$. 

Let \textit{(ii)} hold that $\beta_{1.2}>0$, $\beta_{2.1}>0$. Then 
\begin{equation} \nonumber
\begin{aligned}
E[l_\tau(Y-&\alpha_{2.1}-\beta_{2.1}X)]E[l_\tau(X-\alpha_{1.2}-\beta_{1.2}Y)] = E[l_\tau(e_{2.1})]E[l_\tau(e_{1.2})] \\
= & \beta_{2.1}\beta_{1.2}E[l_{1-\tau}(X+\frac{\alpha_{2.1}}{\beta_{2.1}}-\frac{1}{\beta_{2.1}}Y)]
E[l_{1-\tau}(Y+\frac{\alpha_{1.2}}{\beta_{1.2}}-\frac{1}{\beta_{1.2}}X)]\\
>&E[l_{1-\tau}(X+\frac{\alpha_{2.1}}{\beta_{2.1}}-\frac{1}{\beta_{2.1}}Y)]E[l_{1-\tau}(Y+\frac{\alpha_{1.2}}{\beta_{1.2}}-\frac{1}{\beta_{1.2}}X)]=E[l_{1-\tau}(\tilde{e}_{2.1})]E[l_{1-\tau}(\tilde{e}_{1.2})]\\
\end{aligned}
\end{equation}
\begin{equation} \label{pf eq}
\begin{aligned}
\ge E[l_{\tau}(X+\frac{\alpha_{2.1}}{\beta_{2.1}}-\frac{1}{\beta_{2.1}}Y)]E[l_{\tau}(Y+\frac{\alpha_{1.2}}{\beta_{1.2}}-\frac{1}{\beta_{1.2}}X)],~~~~~~~~~~~~~~~~~~~~~~~~~~~~~~~
\end{aligned}
\end{equation}
because of $(2\tau-1)\Delta_\tau \ge 0$, where $\tilde{e}_{2.1}=-e_{2.1}/\beta_{2.1}$ and $\tilde{e}_{1.2}=-e_{1.2}/\beta_{1.2}$. 
Then, the assumption is a contradiction because $(\alpha_{2.1},\beta_{2.1})=argmin_{(\alpha,\beta)}L_\tau^{X,Y}(\alpha,\beta)$ and $(\alpha_{1.2},\beta_{1.2})=argmin_{(\alpha,\beta)}L_\tau^{Y,X}(\alpha,\beta)$. Therefore, $\beta_{2.1}\beta_{1.2} \le 1.$

We complete the proof by showing the inequality in (\ref{pf eq}).
It is easy to show 
$$l_{1-\tau}(e)=l_\tau(\frac{\tau}{1-\tau}e) = \frac{\tau}{1-\tau}l_\tau(e),~~ \text{if}~ e<0;~~l_{1-\tau}(e)=l_\tau(\frac{1-\tau}{\tau}e) = \frac{1-\tau}{\tau}l_\tau(e),~~ \text{if}~ e\ge0.$$
Therefore,   
\begin{equation} \nonumber
\begin{aligned}E[l_{1-\tau}(\tilde{e}_{1.2})] & E[l_{1-\tau}(\tilde{e}_{2.1})] \\ = \{&(\frac{\tau}{1-\tau})^2E[l_\tau(\tilde{e}_{2.1}^-)]E[l_\tau(\tilde{e}_{1.2}^-)]+
(\frac{1-\tau}{\tau})^2E[l_\tau(\tilde{e}_{2.1}^+)]E[l_\tau(\tilde{e}_{1.2}^+)]\}\\
& +\{E[l_\tau(\tilde{e}_{2.1}^-)]E[l_\tau(\tilde{e}_{1.2}^+)]+E[l_\tau(\tilde{e}_{2.1}^+)]E[l_\tau(\tilde{e}_{1.2}^-)]\} 
= h_1+h_2 \ge E[l_\tau(\tilde{e}_{2.1})]E[l_\tau(\tilde{e}_{1.2})]
\end{aligned}
\end{equation}
if $h_1 \ge E[l_\tau(\tilde{e}_{2.1}^-)]E[l_\tau(\tilde{e}_{1.2}^-)]+E[l_\tau(\tilde{e}_{2.1}^+)]E[l_\tau(\tilde{e}_{1.2}^+)]=h_3$. We have (\ref{pf eq}) because, with $E[l_\tau(\tilde{e}^-)]=(\tau-1)E[\tilde{e}^-]$, $E[l_\tau(\tilde{e}^+)]=\tau E[\tilde{e}^+]$ and the condition of Theorem \ref{thm prop}, we have $h_1-h_3=(2\tau-1)(E[\tilde{e}_{1.2}^-]E[\tilde{e}_{2.1}^-] -E[\tilde{e}_{1.2}^+]E[\tilde{e}_{2.1}^+])
=(2\tau-1)([E[e_{1.2}^-]E[e_{2.1}^-]-E[{e}_{1.2}^+]E[{e}_{2.1}^+])/(\beta_{1.2}\beta_{2.1})=\frac{(2\tau-1)}{\beta_{1.2}\beta_{2.1}}\Delta_\tau\ge0$.

\vspace{3mm}
\noindent\textbf{Proof of Theorem \ref{thm norm}.}
Let $\tilde{F}_Y$ and $\tilde{F}_X$ be the distribution functions of $Y-E[Y|X]$ given $X$, $X-E[X|Y]$ given $Y$, respectively. 
Then, $\tilde{F}_Y$ is free from $X$ and $\tilde{F}_X$ is free from $Y$. By the assumption, $E[Y|X] = \alpha_{2.1}^L+\beta_{2.1}^LX$ and $E[X|Y]=\alpha_{1.2}^L+\beta_{1.2}^LY$ for some $(\alpha_{2.1}^L,\beta_{2.1}^L, \alpha_{1.2}^L, \beta_{1.2}^L)$. Then, $$Q_Y(\tau|X)=\alpha_{2.1}^L+\beta_{2.1}^LX+\tilde{F}_Y^{-1}(\tau),~~Q_X(\tau|Y)=\alpha_{1.2}^L+\beta_{1.2}^LY+\tilde{F}_X^{-1}(\tau).$$ 
By Theorem \ref{coef}, $\beta_{2.1}(\tau) = \beta_{2.1}^L$, $\beta_{1.2}(\tau)=\beta_{1.2}^L$ and hence $\rho_\tau=\rho.~\Box$ 
  \vspace{3mm}

\noindent\textbf{Proof of Theorem \ref{thm TD=0}.}
For random vector $(X,Y)$ satisfying the conditions of Theorem \ref{thm norm}, 
we have $\rho_\tau=\rho$ for all $\tau$. We therefore have
$\rho_\tau=\rho_{0.5}={\color{black}\rho_{1-\tau}}$ for all $\tau$ and hence
$\rho_\tau^D=\rho_\tau-\rho_{0.5}=0$ and $\rho_\tau^A=\rho_\tau-\rho_{1-\tau}=0$.

\vspace{3mm}
   
\noindent\textbf{Proof of Lemma \ref{lem beta}.}
Let $\delta$ be block matrix $\delta = [\delta_1' ~|~\delta_2']'$ and $\delta_1$, $\delta_2$ $\in R^2$. 
We define
$$ z_{1n}(\delta_1) =\sum_{i=1}^n l_\tau(u_{1i}-\delta_1'X_i^* / \sqrt{n})-l_\tau(u_{1i}),~~u_{1i} = Y_i- \theta_{2.1}'(\tau)X_i^*,$$
$$ z_{2n}(\delta_2) =\sum_{i=1}^n l_\tau(u_{2i}-\delta_2'Y_i^* / \sqrt{n})-l_\tau(u_{2i}),~~u_{2i} = X_i- \theta_{1.2}'(\tau)Y_i^*.$$
Note that $z_{1n}(\delta_1)$ and $z_{2n} (\delta_2)$ are convex and are uniquely minimized at 
$\hat{\delta}_{1n} = \sqrt{n} (\hat{\theta}_{2.1}(\tau)-\theta_{2.1}(\tau))$ and $\hat{\delta}_{2n} = \sqrt{n} (\hat{\theta}_{1.2}(\tau)-\theta_{1.2}(\tau)),$ respectively.  
We show 
\begin{equation}\label{eq Z}
Z_n(\delta) = \begin{pmatrix} z_{1n}(\delta_1) \\ z_{2n}(\delta_2)
\end{pmatrix} \goesd \begin{pmatrix} z_{10}(\delta_1) \\ z_{20}(\delta_2)
\end{pmatrix}=Z_0(\delta) 
\end{equation} 
for some $Z_0(\delta)$ and that $Z_0(\delta)$ is uniquely minimized by $\delta_0$ whose distribution is (\ref{eq lem beta}). Then we get the desired result from uniqueness of the minimizers
$$\sqrt{n}(\hat{\Theta}(\tau)-\Theta(\tau)) = \hat{\delta}_n = \begin{pmatrix}
argmin_{\delta_1} z_{1n}(\delta_1) \\ argmin_{\delta_2} z_{2n}(\delta_2) 
\end{pmatrix},~~ \delta_0 = 
\begin{pmatrix}
argmin_{\delta_1} z_{10}(\delta_1) \\ argmin_{\delta_2} z_{20}(\delta_2)
\end{pmatrix}.$$
It remains to prove (\ref{eq Z}).
Let $\alpha_{2.1} = \alpha_{2.1}(\tau)$, $\beta_{2.1}=\beta_{2.1}(\tau)$, $\alpha_{1.2}=\alpha_{1.2}(\tau)$, $\beta_{1.2} = \beta_{1.2}(\tau)$, $\theta_{2.1}=(\alpha_{2.1},\beta_{2.1})'$, $\theta_{1.2}=(\alpha_{1.2},\beta_{1.2})'.$
Applying the Knight (1998)'s identity, we can write 
$$Z_n(\delta) = A_n(\delta) + B_n(\delta),$$
where
\begin{equation} \nonumber
\begin{aligned}
A_n(\delta) = & -\frac{1}{\sqrt{n}}\sum_{i=1}^n \begin{pmatrix}
\delta_1'X_i^*(\tau-I(Y_i-\theta_{2.1}'X_i^*<0)), &
\delta_2'Y_i^*(\tau-I(X_i-\theta_{1.2}'Y_i^*<0))
\end{pmatrix}
 \\
=& -\frac{1}{\sqrt{n}}\sum_{i=1}^n \delta'D_\tau(X_i, Y_i, \theta_{2.1},\theta_{1.2})
\end{aligned}
\end{equation}
and 
\begin{equation}
\begin{aligned}
B_n(\delta) =& \sum_{i=1}^n \begin{pmatrix}
\int^{\delta_1'X_i^*/\sqrt{n}}_0 (I(u_{1i} \le s) - I(u_{1i} \le 0)) ds, &
\int^{\delta_2'Y_i^*/\sqrt{n}}_0 (I(u_{2i} \le s) - I(u_{2i} \le 0)) ds 
\end{pmatrix}'\\
=&\sum_{i=1}^n (B_{1ni}(\delta_1), B_{2ni}(\delta_2))'=(B_{1n}(\delta_1),B_{2n}(\delta_2))'.
\end{aligned}
\end{equation}
We will show
\begin{equation}\label{eq A}
A_n(\delta) \goesd -\delta' W,~~W=[W_1' | W_2']' \sim N_4(0,H_{\tau\tau}(\theta_{2.1},\theta_{1.2})),
\end{equation}
\begin{equation}\label{eq B}
B_n(\delta) \goesp \frac{1}{2}\delta'M(\theta_{2.1},\theta_{1.2})\delta.
\end{equation}
We then have (\ref{eq Z}) with $Z_0(\delta) =-\delta'W+\frac{1}{2}\delta'M(\theta_{2.1},  \theta_{1.2}) \delta$ which is minimized by $\delta_0 = M^{-1}(\theta_{2.1}, \theta_{1.2})W$ having the distribution (\ref{eq lem beta}).

We prove (\ref{eq A}). Let condition A1(i) for iid-ness of $(X_i,Y_i)$ hold. Then $E[D_{\tau i}|F_{i-1}]=E[D_{\tau i}]=0$ by the following argument. 
Noting that $\alpha_{2.1}$ minimizes $L_\tau^{X,Y}(\alpha,\beta_{2.1})=E[l_\tau(Y-\alpha-\beta_{2.1}X)]$, $\alpha_{2.1}$ is the $\tau$-quantile of $Y-\beta_{2.1}X$. Therefore, $E[\tau-I(Y-\alpha_{2.1}-\beta_{2.1}X<0)]=0$. 
Note that $\beta_{2.1}$ minimizes $L_\tau^{X,Y}(\alpha_{2.1},\beta)=E[l_\tau(Y-\alpha_{2.1}-\beta X)]=f(\beta),$ which is differentiable with respect to $\beta$. Therefore, $\frac{\partial f(\beta_{2.1})}{\partial \beta}=0$. Note that, the function $g:(X,Y;\beta)\rightarrow l_\tau(Y-\alpha_{2.1}-\beta X)$ is Lebesgue-integrable with respect to the probability measure of the joint distribution of $(X,Y)$, and for almost all $(X,Y)$, the derivative $g_\beta=\frac{\partial g}{\partial\beta}$ exists for almost all $\beta$. Note that $|g_\beta|\le \max (\tau,1-\tau)|X|,$ a.s. which is integrable. Therefore, the Leibniz's rule is applicable to change the order of integration and derivation as in $0=\frac{\partial f(\beta_{2.1})}{\partial \beta}=E[\frac{\partial}{\partial \beta}l_\tau(Y-\alpha_{2.1}-\beta_{2.1}X)] =E[X(\tau-I(Y-\alpha_{2.1}-\beta_{2.1}X<0))]=0$. Therefore, $E[X_i^*(\tau-I(Y_i-\alpha_{2.1}-\beta_{2.1}X_i<0))]=0$, and similarly $E[Y_i^*(\tau-I(X_i-\alpha_{1.2}-\beta_{1.2}Y_i<0))]=0$, arriving at $E[D_{\tau i}]=0$. Since $(X_i,Y_i),i=1,\cdots,n$ are iid, (\ref{cond A1 eq}) holds automatically. 

Let condition A1(ii) for possibly non-iid sample hold. Then $E[D_{\tau i}|F_{i-1}]=0$.
Therefore, under condition A1(i) or A1(ii) with conditions A2 and A3, applying the martingale central limit theorem to the martingale $\sum_{i=1}^n D_{\tau i}$ with (\ref{cond A1 eq}), we get (\ref{eq A}).

It remains to prove (\ref{eq B}). Observe that
\begin{equation}\nonumber
\begin{aligned}
E[B_{1n}(\delta_1)] =& \sum E[B_{1ni}(\delta_1)|F_{i-1}] = \sum E[E[B_{1ni}(\delta_1)|X_i^*]|F_{i-1}]]\\
=& \sum E[\int_{0}^{\delta_1'X_i^*/\sqrt{n}} F_{Y_i\cdot X_i}(\theta_{2.1}'X_i^*+s)-F_{Y_i\cdot X_i}(\theta_{2.1}'X_i^*)ds|F_{i-1}]\\
=& \sum E[\frac{1}{n}\int_0^{\delta_1'X_i^*} \sqrt{n} \{F_{Y_i\cdot X_i}(\theta_{2.1}'X_i^*+t/\sqrt{n})-F_{Y_i\cdot X_i}(\theta_{2.1}'X_i^*)\}dt|F_{i-1}]\\
=& {\color{black}\sum E[\frac{1}{n}\int_0^{\delta_1'X_i^*}f_{Y_i\cdot X_i}(\theta_{2.1}'X_i^*)t dt|F_{i-1}]} +o(1),~\text{ by condition A2, }\\
=&{\color{black} \frac{1}{2n}} \sum E[f_{Y_i\cdot X_i}(\theta_{2.1}'X_i^*)\delta_1'X_i^*X_i^{*'}\delta_1|F_{i-1}] +o(1) = {\color{black}\frac{1}{2n}}\sum \delta_1' E[R(f_{Y_i\cdot X_i}, X_i, \theta_{2.1})|F_{i-1}]\delta_1+o(1).\\
\end{aligned}
\end{equation}
\begin{equation} \label{eq B3}
\rightarrow  ~\frac{1}{2}\delta_1'\bar{R}_{2.1}(\theta_{2.1})\delta_1
\end{equation}

Note that $[B_{1ni}(\delta_1)]^2 = \int_0^{\delta_1'X_i^*}[I(u_{1i}\le s)-I(u_{1i}\le 0)]dsB_{1ni}(\delta_1)$ and $|B_{1ni}(\delta_1)|\le 2\max_i|\delta_1'X_i^*| |B_{1ni}(\delta_1)|$. By (\ref{eq B3}), essentially for large $n$, $E[B_{1n}(\delta_1)]=\sum E[B_{1ni}(\delta_1)]<\infty$.  Therefore, by (\ref{eq B3}) and  condition A3 (ii),
\begin{equation}\nonumber
\begin{aligned}
Var[B_{1n}(\delta_1)]  \le  \frac{2}{\sqrt{n}}E[\max_i| \delta_1'X_i^*|]E[B_{1n}(\delta_1)] \rightarrow 0.
\end{aligned}
\end{equation}
Therefore, $B_{1n}(\delta_1) \goesp \frac{1}{2} \delta_1'\bar{R}_{2.1}(\theta_{2.1})\delta_1$ and similarly $B_{2n}(\delta_2) \goesp \frac{1}{2} \delta_2'\bar{R}_{1.2}(\theta_{1.2})\delta_2$, arriving at (\ref{eq A}).

\vspace{3mm}

\noindent\textbf{Proof of Corollary \ref{cor beta}.}
By Lemma \ref{lem beta}, we have $\hat{\beta}_{2.1}(\tau)\hat{\beta}_{1.2}(\tau) \goesp \beta_{2.1}(\tau)\beta_{1.2}(\tau)$ and by Theorem \ref{thm welldef}, $\beta_{2.1}(\tau)\beta_{1.2}(\tau)\ge 0, $ for all $\tau$. We therefore get the desired result.

	\vspace{3mm}

\noindent\textbf{Proof of Theorem \ref{thm rhoq}.}
From Lemma \ref{lem beta}, the result is derived easily by applying the multivariate $\delta$-method.

\vspace{3mm}

\noindent\textbf{Proof of Lemma \ref{lem asym two tau}.}
From {\color{black}Lemma} \ref{lem beta}, the result can be obtained by extending the asymptotic normality of four parameter estimators $\hat{\Theta}(\tau)$ to eight parameter estimators {\color{black}$(\hat{\Theta}'(\tau_1), \hat{\Theta}'(\tau_2))'$}. We redefine block matrix {\color{black}$\delta(\tau) = [\delta_1'(\tau)| \delta_2'(\tau)]',~\delta_1(\tau),\delta_2(\tau)\in R^2$}. By Lemma \ref{lem beta}, it suffices to show 
$$ \begin{pmatrix} Z_n(\delta(\tau_1)) \\ Z_n(\delta(\tau_2)) \end{pmatrix} = \begin{pmatrix} z_{1n}(\delta_1(\tau_1)) \\ z_{2n}(\delta_2(\tau_1)) \\z_{1n}(\delta_1(\tau_2)) \\ z_{2n}(\delta_2(\tau_2))\end{pmatrix}
\goesd
\begin{pmatrix} z_{10}(\delta_1(\tau_1)) \\ z_{20}(\delta_2(\tau_1)) \\z_{10}(\delta_1(\tau_2)) \\ z_{20}(\delta_2(\tau_2))\end{pmatrix} = \begin{pmatrix} Z_0(\delta(\tau_1)) \\ Z_0(\delta(\tau_2)) \end{pmatrix}. $$
Applying the Knight (1998)'s identity, we can write 
$$\begin{pmatrix} Z_n(\delta(\tau_1)) \\ Z_n(\delta(\tau_2)) \end{pmatrix}=\begin{pmatrix} A_n(\delta(\tau_1)) \\ A_n(\delta(\tau_2)) \end{pmatrix} + \begin{pmatrix} B_n(\delta(\tau_1)) \\ B_n(\delta(\tau_2)) \end{pmatrix},$$
and we get the desired result by showing 
\begin{equation}\label{eq A2}
\begin{aligned}
~~~~\begin{pmatrix} A_n(\delta(\tau_1)) \\ A_n(\delta(\tau_2)) \end{pmatrix}  \goesd{\color{black} [\delta'(\tau_1)|\delta'(\tau_2)]}W_8,~~W_8\sim N_8(0,V_0),
\end{aligned}
\end{equation}
\begin{equation}\label{eq B2}
\begin{aligned}
\begin{pmatrix} B_n(\delta(\tau_1)) \\ B_n(\delta(\tau_2)) \end{pmatrix} \goesp \frac{1}{2}  {\color{black} [\delta'(\tau_1)|\delta'(\tau_2)]} V_1 {\color{black} [\delta'(\tau_1)|\delta'(\tau_2)]'}.
 \end{aligned}
 \end{equation}
Proofs of (\ref{eq A2}) and (\ref{eq B2}) are the same as those for (\ref{eq A}) and (\ref{eq B}).

\vspace{3mm}

\noindent\textbf{Proof of Theorem \ref{thm asym two tau}.}
From Lemma \ref{lem asym two tau}, the result is derived easily by applying the multivariate $\delta$-method. 

\vspace{3mm}

\noindent\textbf{Proof of Corollary \ref{test TD}.}  
Theorem \ref{thm asym two tau} with $\tau_1=\tau$ and $\tau_2=0.5$ gives the result. 

\vspace{3mm}

\noindent\textbf{Proof of Corollary \ref{test TA}.}  
Theorem \ref{thm asym two tau} with $\tau_1=\tau$ and $\tau_2=1-\tau$ gives the result.

\vspace{5mm}
\section*{ Acknowledgements}  This study was supported by a grant  from
the National Research Foundation of Korea (2016R1A2B4008780).

\vspace{5mm}

\section*{References}

{\small
	\refmark Abreveya, J. 2001. The effects of demographics and maternal behavior on the distribution of birth outcomes, Empirical Economics, 26, 247-257.
	
	 \refmark Adrian, T., Brunnermeier, M. K. 2016. CoVaR, American Economic Review, 106, 1705-1741.
	 
	 \refmark Bashtannyk, D. M.,  Hyndman, R. J. 2001. Bandwidth selection for kernel conditional density estimation, Computational Statistics \& Data Analysis, 36, 279-298.
	 
	 \refmark Bofinger, E. 1975. Estimation of a density function using order statistics, Australian Journal of Statistics, 17, 1-7. 
	 
	  \refmark Diebold, F. X., Yilmaz, K. 2012. Better to give than to receive: Predictive direction measurement of volatility spillovers, International Journal of Forecasting, 28, 57-66. 
	 
	 \refmark Girardi, G., Ergun, T. 2013. Systemic risk measurement: Multivariate GARCH estimation of CoVaR, Journal of Banking \& Finance, 37, 3169-3180.
	 
	  \refmark Hendricks, W., Koenker, R. 1991. Hierarchical spline models for conditional quantiles and the demand for electricity, Journal of the American Statistical Association, 87, 58-68. 
 
	 \refmark Hyndman, J. R, Bashtannyk, D. M., Grunwald, G. K. 1996. Estimating and visualizing conditional densities, Journal of Computational and Graphical Statistics, 5, 315-336. 
	 
	  \refmark Joe, H., Li, H., Nikoloulopoulos, A. K. 2010. Tail dependence functions and vine copulas, Journal of Multivariate Analysis, 101, 252-270.
	  
	  \refmark Knight, K. 1998. Limiting distributions for $L_1$ regression estimators under general conditions, Annals of Statistics, 26, 755-770.
	  
	  \refmark Kocherginsky, M., He, X., Mu, Y. 2005. Practical confidence intervals for regression quantiles, Journal of Computational and Graphical Statistics, 14, 41-55.
	  
	  \refmark Koenker, R. 2005. Quantile Regression, New York, Cambridge University Press.

	 \refmark Kollo, T., Pettere, G., Valge, M. 2017. Tail dependence of skew t-copulas, Communications in Statistics - Simulation and Computation, 46, 1024-1034.

	 \refmark Li, G., Li, Y., Tsai, C-L. 2015. Quantile correlations and quantile autoregressive modeling, Journal of the American Statistical Association, 110, 246-261.

	 \refmark Meng, L., Shen, Y. 2014. On the relationship of soil moisture and extreme temperatures in east China,  Earth Interactions, 18, 1-20.
	 
	 \refmark Nikoloulopoulos, A. K., Joe, H., Li, H. 2012. Vine copulas with asymmetric tail dependence and applications to financial return data, Computational Statistics \& Data Analysis, 56, 3659-3673.

	 \refmark Sayegh, A. S., Munir, S., Habeebullah, T. M. 2014. Comparing the performance of statistical models for predicting $PM_{10}$ concentration, Aersol and Air Quality Research, 14, 653-665.
	 
	\refmark Villarini, G., Smith, J. A., Baeck, M. L., Vitolo, R., Stephenson, D. B., Krajewski, W. F. 2011. On the frequency of heavy rainfall for the Midwest of the United States, Jounral of Hydrology, 400, 103-120. 
}

\end{document}